\documentclass[preprint,showpacs,preprintnumbers,amsmath,amssymb,floatfix]{revtex4-1}
\newcommand{\newwidth}{0.675\textwidth}
\newcommand{\newheight}{0.45\textwidth}

\usepackage{slashbox}
\usepackage{graphicx}
\usepackage{dcolumn}
\usepackage{bm,verbatim}
\usepackage{sidecap}

\begin{document}

\title{Decoherence and lead induced inter-dot coupling  
in nonequilibrium electron transport through interacting quantum dots: 
A hierarchical quantum master equation approach} 

\author{R.\ H\"artle$^{1,}$\footnote{\emph{Present Address:} 
Institut f\"ur theoretische Physik, Georg-August-Universit\"at G\"ottingen, Friedrich-Hund-Platz 1, D-37077 G\"ottingen, Germany}}
\author{G.\ Cohen$^{2}$}
\author{D.\ R.\ Reichman$^{2}$}
\author{A.\ J.\ Millis$^{1}$}
\affiliation{
$^1$ Department of Physics, Columbia University, New York, NY 10027, USA. \\
$^2$ Department of Chemistry, Columbia University, New York, NY 10027, USA. 
}

\date{\today}

\begin{abstract}
The interplay between interference effects and electron-electron interactions in electron 
transport through an interacting double quantum dot system is investigated using a hierarchical 
quantum master equation approach which becomes exact if carried to infinite order and 
converges well if the temperature is not too low. Decoherence due to electron-electron interactions 
is found to give rise to pronounced negative differential resistance, enhanced broadening of 
structures in current-voltage characteristics and an inversion of the electronic population. 
Dependence on gate voltage is shown to be a useful method of distinguishing decoherence-induced 
phenomena from effects induced by other mechanisms such as the presence of a blocking state. 
Comparison of results obtained by the hierarchical quantum master equation approach to those 
obtained from the Born-Markov approximation to the Nakajima-Zwanzig equation and from the non-crossing 
approximation to the nonequilibrium Green's function reveals the importance of an inter-dot coupling 
that originates from the energy dependence of the conduction bands in the leads and the need for a systematic perturbative expansion. 
\end{abstract}


\maketitle

\section{Introduction}

Electron transport through nanoelectronic devices involves fundamentally important principles which 
often result in interesting technological applications \cite{Kastner2000,cuevasscheer2010}. Resonant 
tunneling diodes exhibit a nonlinear current-voltage response, in particular negative differential 
resistance, due to the quantization of the respective energy levels \cite{Tsu1974,Davies93,Mizuta1995}.  
Quantum interference phenomena may be used to control the current flow in three-terminal nanoscale 
transistors \cite{Stafford2007,Saha2010} and single-molecule junctions \cite{Ballmann2012}.  
Interaction-driven phenomena such as, for example, static or dynamical Coulomb blockade \cite{Kastner2000,Kiesslich2007,Brun2012}  
or the Kondo effect \cite{Cronenwett1998,Goldhaber1998,Liang02} also occur. 
While these phenomena have been studied separately, the interplay between interference phenomena, 
level quantization and  electron-electron interactions in nanoelectronic devices has been less studied 
and is an important open problem.

Interference effects may arise from a spatial separation of tunneling pathways. As one of many examples 
one may mention quantum dot arrays  set up as Aharonov-Bohm interferometers \cite{Holleitner2001,Iye2001,Holleitner2002,Wegscheider2007}.  
Quasidegenerate energy levels of a single quantum dot may also be understood as separate tunneling pathways 
and the respective conduction properties interpreted in terms of quantum 
interference \cite{Kalyanaraman2002,Ernzerhof2005,Solomon2008b,Brisker2010,Markussen2010,Lambert2011,Brisker2012,Hartle2011,Ballmann2012,Hartle2012}.  
Both scenarios may be described by Hamiltonians that are nearly identical.

Interference may be strongly affected by electron-electron interactions or coupling between electrons 
and vibrational modes. Decoherence phenomena arising from electron-vibrational mode coupling in quantum dot systems \cite{Marquardt2003,Ueda2010}
or single-molecule junctions \cite{Hartle2011b,Hartle2012} have been studied; one striking result is a pronounced temperature 
dependence of the current \cite{Ballmann2012,Hartle2012,Hartle2013}. In quantum dot arrays, interference effects are suppressed by 
spin-flip processes \cite{Gefen2001,Iye2001,Gefen2002}. A similar effect is observed in InSb nanowires, where 
interaction- or correlation-induced resonances occur \cite{Meden2006,Lee2007,Schiller2009,Nilsson2010}.

In this article, we investigate the interrelation between interference effects and electron-electron 
interactions in a nonequilibrium nanoelectronic device modeled as a spinless two-orbital Anderson impurity  
coupled to two leads, which may be maintained at different chemical potentials. The two-orbital Anderson model  
is perhaps the simplest model where the interplay of interference effects and decoherence phenomena due to 
electron-electron interactions can be theoretically studied.  It has been considered before  by a number of 
authors 
\cite{Hofstatter2001,Gefen2002,Marquardt2003,Meden2006,Lee2007,Wunsch2005,Hettler2007,Kashcheyevs2007,Kiesslich2007,Kubo2008,Schiller2009,Trocha2009,Nilsson2010,RouraBas2011,Karlstrom2011,Trocha2012,Segal2012}  
and may be physically realized in a device where the  spin degeneracy is lifted by an  external magnetic field or  
spin-polarized leads. Despite its deceptively simple structure, the spinless Anderson impurity model manifests a rich variety of physical phenomena, 
including orbital/pseudospin-Kondo physics \cite{Hofstatter2001,Kashcheyevs2007,Kubo2008,Trocha2012}, 
population inversion \cite{Lee2007,Goldstein2010,Karlstrom2011},  
negative differential resistance (NDR) \cite{Wunsch2005,Hettler2007,Trocha2009,RouraBas2010}, Fano-line 
shapes \cite{Goldstein2007,Wegscheider2007}, interaction-induced level repulsion \cite{Hofstatter2001,Wunsch2005} 
and resonances \cite{Meden2006,Lee2007,Schiller2009,Nilsson2010}.

\begin{figure}
\begin{center}
\includegraphics[width=0.5\textwidth]{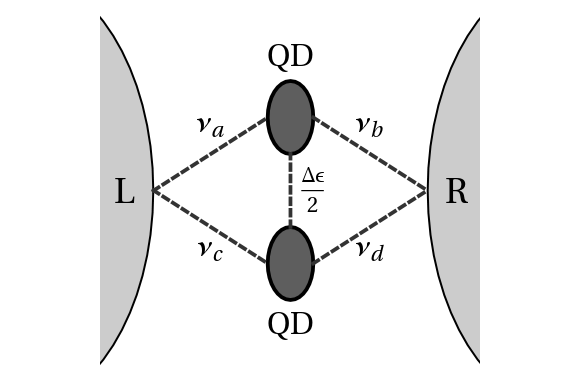}
\end{center}
\caption{\label{LinConduct} 
Schematic representation of the nanoelectronic device that is considered in the text. 
It consists of two quantum dots (QD) that are coupled with each other and to a left (L) and a right (R) electrode. 
In this work, the inter-dot coupling is assumed to be non-zero and smaller than the coupling to the leads, 
$0<\Delta\epsilon/2\ll\nu_{x}$, $x\in\{a,b,c,d\}$. Two cases of particular interest are coupling in 
a serial ($\nu_{a}=\nu_{d}\neq0$, $\nu_{b}=\nu_{c}=0$) and a branched form ($\nu_{a}=\nu_{b}\neq0$, $\nu_{c}=\nu_{d}=0$).
}
\end{figure}

While different realizations of this model are possible (cf.\ Fig.\ \ref{LinConduct}), 
we focus in the following on two complementary cases, which show the most relevant and 
pronounced interference effects. This includes scenarios, where the two dots (or localized orbitals)   
are coupled in either a serial or a branched form, including only a weak coupling between the 
two dots (see Fig.\ \ref{LinConduct}). Thereby, the corresponding eigenstates are coupled to the electrodes 
in the same way as two quantum dots in an Aharonov-Bohm interferometer \cite{Holleitner2001,Iye2001,Holleitner2002,Wegscheider2007} 
such that the two realizations may also correspond to the two extreme cases where 
the magnetic flux is $\pi$ and zero, respectively (in units of the flux quantum and modulo multiples of $2\pi$).

In this work, we identify the finite bandwidth of the leads as an important but previously poorly studied variable. 
Our detailed results show that it can induce a coupling between the eigenstates of the serial and branched realization 
or, equivalently, between the dots of a corresponding AB interferometer. This coupling gives rise to 
population inversion and a strong renormalization of the corresponding signatures in the transport characteristics, 
in particular an enhanced broadening. It is similar to Ruderman-Kittel-Kasuya-Yosida (RKKY) (spin-spin) 
interactions in solids \cite{Hewson93,Craig2004,Sasaki06,Kulkarni2010,Tutuc2011}  
and has also been referred to in the literature as an indirect coupling \cite{Kubo2008,Trocha2012}, where, however, it 
is associated with the fact that the two eigenstates/dots are coupled to the same leads rather than the energy dependence of 
the respective conduction bands.

Electron transport through such nanoelectronic devices has been studied using approximate methods like 
second \cite{Hettler2002,Hettler2003,Datta2007,Darau2009,Hartle09,Donarini2010,Hartle2010b} 
and higher order \cite{Pedersen05,Hettler2007,Leijnse09,Esposito2010} 
master equation methodologies, real-time diagrammatic techniques \cite{Koenig96,Koenig1996,Tews2004,Wunsch2005}, 
nonequilibrium Green's function methods \cite{Meir1991,Groshev,Fransson2005,Thygesen2008,Stafford09,Hartle09,Galperin0908,RouraBas2011} 
and (nonequilibrium) scattering state approaches \cite{LeHur2010}. 
Numerically exact schemes based on time-convolutionless master equations \cite{Timm2010}, 
numerical \cite{Anders2005,Anders2008,Korb2007} and functional \cite{Meden2005,Meden2006} renormalization group theory \cite{Schiller2009}, 
density matrix renormalization group methods \cite{Schmitteckert04,Novotny2009,HeidrichMeisner2009}, 
flow equation approaches \cite{Kehrein2005,Hackl10,Hackl2010}, 
iterative \cite{Thorwart2008,Segal2010,Huetzen2012} and stochastic \cite{Werner2009,Schiro2010,Han2010,Muhlbacher2011,Gull2011} 
diagrammatic methods and wave-function propagation algorithms \cite{Wang09,Thoss2013} have also been used. 
Additionally, numerically exact reduced dynamics techniques, which exist for population dynamics \cite{Cohen2011} and 
transport properties, \cite{Cohen2013b} have been applied to both the stochastic diagrammatic methods \cite{Cohen2013}  
and the wave function propagation schemes \cite{Wilner2013}.

Our studies are based on the hierarchical quantum master equation (HQME) approach introduced by Jin \emph{et al.}\ \cite{Jin2008} 
and modified by us in several ways, of which  the most important is the use of a  truncation scheme different from that 
proposed in Ref.~\onlinecite{Shi2009}.  The HQME method is in effect a perturbative expansion in powers of the dot-lead hybridization 
divided by the temperature; if carried to infinite order it is exact and if the temperature is not too low, convergence 
can be verified numerically. The advantages of the HQME method are that it is time-convolutionless, non-perturbative in the 
electron-electron interaction and gives numerically exact access to the  steady state properties of  nanoelectronic devices even in situations (such as those involving  
quasidegenerate levels \cite{Timm2010}) where electronic relaxation time scales can become relatively long.

For comparison, we also employ two approximate methods: the  Born-Markov (BM) master equation method 
\cite{May02,Hettler2002,Mitra04,Lehmann04,Harbola2006,Hartle09,Hartle2010b} and a non-crossing approximation (NCA) calculation of the nonequilibrium 
Green's function
\cite{Keiter1970,Keiter1971,Grewe1981,Kuramoto1983,Coleman1984,Bickers1987b,Bickers1987,Nordlander1999,Eckstein2010,Gull2011}.  
The comparison reveals the importance of a systematic (hybridization) expansion. In addition, it allows us to assess the role of 
renormalization and inter-state/inter-dot coupling effects arising from finite lead bandwidths (which is typically neglected) and the 
differences arising from different choices of lead density of states. Note that, only recently, the HQME framework 
was compared to master equation and nonequilibrium Greensfunction approaches in Ref.\ \cite{Popescu2013}, where 
simple (\emph{i.e.} non-interacting) time-dependent transport problems were studied.

The article is organized as follows. In Sec.\ \ref{theory}, we outline the theoretical methodology, including  a brief 
description of the spinless Anderson impurity model (Sec.\ \ref{modham}), the HQME approach (Sec.\ \ref{heomtheory}),  
the BM master equation scheme (Sec.\ \ref{bornmarkovtheory}) and the NCA based Green's function scheme (Sec.\ \ref{ncatheory}).  
Our results are presented in Sec.\ \ref{results},  where we discuss the transport characteristics of the 
serial (Sec.\ \ref{dessec}) and the branched (Sec.\ \ref{consec}) realizations of the spinless 
two-orbital Anderson model. On the single-particle level the parallel realization exhibits constructive 
interference. Here, the effect of (repulsive) electron-electron interactions can be readily understood as a blocking of 
transport channels in this system. In contrast, the serial conduction case shows pronounced destructive 
interference and decoherence induced phenomena 
which are analyzed in detail in Secs.\ \ref{desnonint} -- \ref{finUres}. Sec.\ \ref{asymsec} then examines the consequences of 
an asymmetric coupling to the electrodes while Sec.\ \ref{gatesec} presents the  gate-voltage dependence of the decoherence 
effects discussed here. A comparison to other systems which show negative differential resistance due to the effect of electron-electron interactions  
is given in Sec.\ \ref{compblock}. In Sec.\ \ref{ncaresults}, we consider the robustness of our conclusions under varying the lead density of states.   
Sec.\ \ref{numerisec} includes a discussion of higher order effects and some numerical aspects.

\section{Theory}
\label{theory}

\subsection{Model Hamiltonian}
\label{modham}

We consider interference effects in electron transport through a nanoelectronic system (S) that supports two quasidegenerate 
electronic states. A schematic is shown in Fig.\ \ref{LinConduct}. Physical realizations include two quantum dots arranged to  
form an Aharonov-Bohm interferometer 
\cite{Holleitner2001,Iye2001,Holleitner2002,Wegscheider2007} or a single nanoscale conductor with an appropriate level 
structure \cite{Kalyanaraman2002,Ernzerhof2005,Solomon2008b,Brisker2010,Markussen2010,Lambert2011,Brisker2012,Hartle2011,Ballmann2012,Hartle2012}. 
We model this situation as a two orbital spinless Anderson model. The corresponding Hamiltonian reads 
\begin{eqnarray}
 H_{\text{S}} &=& \epsilon_{1} d_{1}^{\dagger}d_{1} + \epsilon_{2} d_{2}^{\dagger}d_{2} 
 + U d_{1}^{\dagger}d_{1} d_{2}^{\dagger}d_{2},
\end{eqnarray}
where  $U$ parametrizes the Coulomb interactions in this system.

The nanoelectronic system S is coupled to  two macroscopic (metal) electrodes (L/R), which provide 
reservoirs of electrons. Departures from equilibrium may be achieved by choosing different chemical potentials or 
temperatures for the two leads.  Each of the electrodes can be represented as a continuum of non-interacting electronic states  
\begin{eqnarray}
\label{hLR}
H_{\text{L/R}} &=& \sum_{k\in\text{L/R}} \epsilon_{k} c_{k}^{\dagger}c_{k}.   
\end{eqnarray} 
The coupling between the system S and the electrodes L and R 
is given by 
\begin{eqnarray}
\label{htun}
H_{\text{tun}} &=& \sum_{k\in\text{L,R};m\in\{1,2\}} ( V_{mk} c_{k}^{\dagger}d_{m} + \text{h.c.} ). 
\end{eqnarray} 
The coupling matrix elements $V_{mk}$ determine the so-called level-width (or coupling density) functions 
\begin{eqnarray}
\Gamma_{K,mn}(\epsilon)=2\pi\sum_{k\in K} 
V_{mk}^{*} V_{nk}\delta(\epsilon-\epsilon_{k})
\end{eqnarray}
with $K\in\{\text{L,R}\}$. 
In the following, we assume these functions to be Lorentzian,   
\begin{eqnarray}
\label{levelwidths}
 \Gamma_{K,mn}(\epsilon) &=& 2\pi \nu_{K,m} \nu_{K,n} \frac{\gamma}{(\epsilon-\mu_{K})^{2}+\gamma^{2}}, 
\end{eqnarray}
with the (band) width parameter $\gamma$, the chemical potential $\mu_{K}$ and  $\nu_{K,m}$ denoting the coupling strength 
between state $m$ and lead $K$.  This form of the level-width functions is advantageous in setting up the basic HQME framework. 
More general level-width  functions can be implemented, using, for example, the Meir-Tannor  parametrization scheme 
\cite{Tannor1999,Welack2006,Jin2008}. The basic physics that is discussed in this paper is not influenced by this choice of the level-width functions 
(cf.\ Sec.\ \ref{ncaresults}).

We choose the zero of energy to be the average of the two lead chemical potentials and define $\Phi$  as the 
bias voltage, so the chemical potentials in the left and the right leads are given by $\mu_{\text{L}}=e\Phi/2$ and 
$\mu_{\text{R}}=-e\Phi/2$. 
This description 
is appropriate to devices with the  structure of the quantum dot array in Fig.\ \ref{LinConduct} and applies to physical 
realizations such as the  Aharonov-Bohm-like setups of Refs.~\onlinecite{Wegscheider2007,Trocha2009,Segal2012} 
as well as to most single-molecule junctions \cite{Kubiak1997,Mujica2002}. Other geometries may require a different 
model for the drop of the bias voltage \cite{Fujisawa1998,Kiesslich2007}. 
The full Hamiltonian $H$ is given by 
\begin{eqnarray}
H &=& H_{\text{S}} + H_{\text{L}} + H_{\text{R}} + H_{\text{tun}}. 
\end{eqnarray}

In the following, we assume the coupling strengths $\nu_{K,m}$ to be energy independent.  While this is not the most 
general case, it provides a sufficient description of the condensed matter systems of main experimental interest. 
The assumption  allows us  to choose an energy independent basis where 
$\Gamma_{\text{L},mn}(\epsilon)\sim \delta_{m,1}\delta_{n,1}$ and $\Gamma_{\text{R},mn}(\epsilon)=0 $ for $m,n>2$ 
and to separate the energy-dependent 
part of $\Gamma_{K,mn}(\epsilon)$, \emph{i.e.}\ $2\pi\gamma/((\epsilon-\mu_{K})^{2}+\gamma^{2})$, from the part 
that depends on  the degrees of freedom of the system S, $\nu_{K,m} \nu_{K,n}$. This allows us to significantly reduce the 
numerical effort in the HQME calculations described next.

\subsection{Hierarchical master equation approach} 
\label{heomtheory}

To calculate the nonequilibrium transport properties of the spinless Anderson model, we use a 
modified version of the HQME method of Jin \emph{et al.}\ \onlinecite{Jin2008,Zheng2009}. 
The approach is based on a representation of the reduced density matrix $\sigma$ in terms of the Feynman-Vernon 
influence functional \cite{Feynman63,Makri95a,Segal2010}. It was originally developed to describe bosonic reservoir 
degrees of freedom \cite{Tanimura1989,Tanimura1990,Tanimura06}, which are important for example in photosynthesis \cite{Kramer2011,Kais2011,Strumpfer2012}. 
To lowest order, it reduces to the non-Markovian density matrix approach of Welack \emph{et al.}\ \cite{Welack2006}. 
The HQME framework is suitable for the present problem because it represents a time-convolutionless master equation approach 
\cite{Timm2010} which allows us to address the long relaxation time scales in the presence of quasidegenerate 
levels and facilitates a non-perturbative description of electron-electron interactions. Moreover, populations and 
coherences of the system are treated on the same footing, which is necessary for the investigation of the complex interplay between 
interactions and quantum interference effects.

The central quantity of the approach is the reduced density matrix $\sigma(t)$.  It is defined by the trace over the leads 
of the total density matrix $\varrho(t)$ 
\begin{eqnarray}
 \sigma(t) &=& \text{Tr}_{\text{L+R}}\left\{ \varrho(t) \right\}.   
\end{eqnarray}
The time dependence of the total density matrix $\varrho(t)$ can be written as  
\begin{eqnarray}
 \varrho(t) &=& U(t,0)  \varrho(0) U^{\dagger}(t,0) 
\end{eqnarray} 
with the time evolution operator 
\begin{eqnarray}
 U(t,0) &=& T\left(\text{e}^{-i \int_{0}^{t}\text{d}\tau \left( H_{\text{tun}}(\tau) + H_{\text{S}} \right)}\right).   
\end{eqnarray}
Here $T$ denotes the time ordering operator. Note that we have written the formalism in an interaction picture, where 
only the lead Hamiltonians $H_{\text{L/R}}$ are used but not the system Hamiltonian $H_{\text{S}}$. 
Thus, we avoid the direct appearance of dynamical phases in the reduced density matrix.

The equation of motion of the reduced density matrix is 
\begin{eqnarray}
 \frac{\text{d}}{\text{d}t} \sigma(t) &=& 
  -i \left[ H_{\text{S}} , \sigma(t) \right] 
 - \sum_{m,s} \left[ d_{m}^{s}, \tilde{\sigma}_{ms}(t) \right] . 
\end{eqnarray}
with 
\begin{eqnarray}
 \sum_{m,s} \left[ d_{m}^{s}, \tilde{\sigma}_{ms}(t) \right] &=& i \text{Tr}_{\text{L+R}}\left\{ \left[ H_{\text{tun}}(t) , \varrho(t) \right]  \right\}. 
\end{eqnarray}
and $s\in\{+,-\}$, $d_{n}^{+}=d_{n}^{\dagger}$ and $d_{n}^{-}=d_{n}$. 
The reduced density matrix can thus be obtained from the operators $\tilde{\sigma}_{ms}(t)$. Writing equations of motion 
for these operators leads to an infinite hierarchy  of equations involving the nested commutators 
$ \left[ H_{\text{tun}}(t) , \left[ H_{\text{tun}}(t) , .... , \varrho(t) \right] \right]$, 
$ \left[ \partial_{t} H_{\text{tun}}(t) , \left[ H_{\text{tun}}(t) , .... , \varrho(t) \right] \right]$ and so on.    
In practical calculations, this set of equations needs to be truncated at some finite level. However, it is not a priori clear which of 
these operators can be neglected; in particular, a systematic method for dropping operators involving  the time 
derivatives $\partial_{t}^{q} H_{\text{tun}}(t)$ with $q\in\mathbb{N}$ seems not to be available.

The HQME approach of  Jin \emph{et al.}\ \cite{Jin2008} solves this problem in  two steps. First, the operators $\tilde{\sigma}_{ms}(t)$ 
are rewritten as 
\begin{eqnarray}
 \tilde{\sigma}_{ms}(t) &=& 
 \sum_{Kn} \int_{0}^{t}\text{d}\tau\, C^{\overline{s}}_{K,mn}(t-\tau) 
 \text{Tr}_{\text{L+R}}\left\{ U(t,\tau) d_{n}^{\overline{s}} U(\tau,0) \varrho(0) U^{\dagger}(t,0)\right\}   \\
 && 
 - \sum_{Kn} \int_{0}^{t}\text{d}\tau\, C^{s,*}_{K,mn}(t-\tau) \text{Tr}_{\text{L+R}}\left\{ U(t,0) \varrho(0) 
U^{\dagger}(\tau,0) 
 d_{n}^{\overline{s}} U^{\dagger}(t,\tau) \right\},  \nonumber
\end{eqnarray} 
using the correlation functions 
\begin{eqnarray}
C^{s}_{K,mn}(t-t') &=& \sum_{k\in K} V^{\overline{s}}_{mk} V_{nk}^{s} \text{Tr}_{K}\left\{ \sigma_{K} c_{k}^{s}(t) c_{k}^{\overline{s}}(t') \right\},  
\end{eqnarray}
where 
\begin{eqnarray}
 \sigma_{K} &=& 
 \frac{1}{\text{Tr}_{K}\left\{ \text{e}^{-\sum_{k\in K}\frac{\epsilon_{k}-\mu_{\text{L/R}}}{k_{\text{B}}T} c_{k}^{\dagger}c_{k}  }\right\} } 
 \text{e}^{-\sum_{k\in K}\frac{\epsilon_{k}-\mu_{\text{L/R}}}{k_{\text{B}}T} c_{k}^{\dagger}c_{k}}, 
\end{eqnarray}
$k_{\text{B}}$ denotes the Boltzmann constant, $T$ the temperature of the electrodes, $\overline{s}=-s$,   
$V_{mk}^{+}=V_{mk}$, $V_{mk}^{-}=V_{mk}^{*}$, $c_{k}^{+}=c_{k}^{\dagger}$ and $c_{k}^{-}=c_{k}$. Note that in deriving 
these expressions it has been assumed that the system is initially in a factorized state, i.e. that  the density matrix 
$\varrho(0)$ at time $t=0$ is given by the product $\sigma(0)\sigma_{\text{L}}\sigma_{\text{R}}$. This choice of the initial 
state is not important in the present context, because we wish to study the steady state properties of the system S.   

Second, the correlation functions $C^{s}_{K,mn}$ are represented by a set of exponential functions,    
\begin{eqnarray}
\label{paramscheme}
C^{s}_{K,mn}(t)  &=&  \frac{\nu_{K,m} \nu_{K,n}}{\gamma} \sum_{p\in\mathbb{N}_{0}} \eta^{s}_{K,p} \text{e}^{- \omega^{s}_{K,p} t} . 
\end{eqnarray} 
which, due to the self similarity of these exponentials with respect to time derivatives, \emph{i.e.} 
$\partial_t \text{exp}(-\omega t) \sim \text{exp}(-\omega t)$, 
can also be used to represent the respective time derivative, $\partial_t C^{s}_{K,mn}(t)$. 
This property is crucial because it will allow us to express the various time derivatives of the hybridization 
operator $H_{\text{tun}}(t)$, which appeared before in the nested commutators 
$ \left[ H_{\text{tun}}(t) , \left[ H_{\text{tun}}(t) , .... , \varrho(t) \right] \right]$, 
$ \left[ \partial_{t} H_{\text{tun}}(t) , \left[ H_{\text{tun}}(t) , .... , \varrho(t) \right] \right]$ and so on,  
in terms of known auxiliary operators rather than an infinite series of unknown ones and, thus, 
to truncate the corresponding equations of motion in a systematic way (vide infra).  
Explicitly,
\begin{eqnarray}
\label{paramscheme3}
 \omega^{\pm}_{K,p} &=& \left\{ \begin{array}{ll}
                                 \gamma \mp i \mu_K, & p=0,\\
                                 \pi k_{\text{B}}T (2p-1) \mp i \mu_{K}, & p\in\mathbb{N}, \\
                                \end{array}\right.
 \\
 \label{paramscheme2}
 \eta^{\pm}_{K,p} &=& \left\{ \begin{array}{ll}
                                 \pi\frac{\gamma}{1+\text{exp}(i\frac{\gamma}{k_{\text{B}}T })}  , & p=0,\\
                                 - 2\pi i k_{\text{B}}T \frac{\gamma^{2}}{ ( \mu_{K}  \mp i \omega_{K,p}^{\pm}  )^{2} +\gamma^{2}}, & p\in\mathbb{N},\\
                                \end{array}\right. 
\end{eqnarray} 
where contour integration has been used to obtain the final result. 
This representation relies on the assumption  that the level-width function $\Gamma_{K,mn}$ can be represented by a set 
of Lorentzians (cf.\ Eq.\ \ref{levelwidths}). The operators $\tilde{\sigma}_{ms}(t)$ can thus be expressed in 
terms of a new class of auxiliary operators 
\begin{eqnarray}
\label{defauxop2} 
 \sigma_{K,n,s,p}(t) &=& \eta^{s}_{K,p} \int_{0}^{t}\text{d}\tau\,  \text{e}^{- \omega^{s}_{K,p} (t-\tau)} 
 \text{Tr}_{\text{L+R}}\left\{ U(t,\tau) d_{n}^{s} U(\tau,0) \varrho(0) U^{\dagger}(t,0)\right\}  \\
 && 
 - \eta^{\overline{s},*}_{K,p} \int_{0}^{t}\text{d}\tau\, \text{e}^{- \omega^{s}_{K,p} (t-\tau)} 
 \text{Tr}_{\text{L+R}}\left\{  U(t,0) \varrho(0) U^{\dagger}(\tau,0) d_{n}^{s}
 U^{\dagger}(t,\tau) \right\}  \nonumber
\end{eqnarray}
as 
\begin{eqnarray}
 \tilde{\sigma}_{ms}(t) &=& \sum_{Knp} \frac{\nu_{K,m} \nu_{K,n}}{\gamma} \sigma_{K,n,\overline{s},p}(t). 
\end{eqnarray}
Note that, in general, 
a decomposition in terms of exponential functions is only possible for non-zero temperatures. At 
$T=0$ correlation functions $C^{s}_{K,mn}$ may, in general, exhibit a $1/t$ dependence, 
which cannot be represented by a set of exponentials.

The equations of motion of the auxiliary operators $\sigma_{K,n,s,p}(t)$ represent the first tier of an infinite 
hierarchy of equations of motion \cite{Jin2008}. The full hierarchy is then written as 
\begin{eqnarray}
\label{hierarcheom}
 \partial_{t} \sigma^{(\alpha)}_{j_{1}..j_{\alpha}}(t) &=&  
 - i \left[ H_{\text{S}} , \sigma^{(\alpha)}_{j_{1}..j_{\alpha}}(t) \right] 
 - \sum_{\beta\in\{1..\alpha\}} \omega_{K_{\beta},p_{\beta}}^{s_{\beta}} \sigma^{(\alpha)}_{j_{1}..j_{\alpha}}(t) \\
 && + \sum_{\beta\in\{1..\alpha\}} (-1)^{\alpha-\beta} \eta_{K_{\beta},p_{\beta}}^{s_{\beta}} d_{m_{\beta}}^{s_{\beta}} \sigma^{(\alpha-1)}_{j_{1}..j_{\alpha}/j_{\beta}}(t) 
 + \sum_{\beta\in\{1..\alpha\}} (-1)^{\beta} \eta_{K_{\beta},p_{\beta}}^{\overline{s}_{\beta},*}  \sigma^{(\alpha-1)}_{j_{1}..j_{\alpha}/j_{\beta}}(t) d_{m_{\beta}}^{s_{\beta}} \nonumber \\
 && - \sum_{j_{\alpha+1},n_{\alpha+1}} \frac{\nu_{K_{\alpha+1},m_{\alpha+1}} \nu_{K_{\alpha+1},n_{\alpha+1}}}{\gamma} \left( d_{n_{\alpha+1}}^{\overline{s}_{\alpha+1}} 
 \sigma^{(\alpha+1)}_{j_{1}..j_{\alpha}j_{\alpha+1}}(t) - (-1)^{\alpha} 
  \sigma^{(\alpha+1)}_{j_{1}..j_{\alpha}j_{\alpha+1}}(t) d_{n_{\alpha+1}}^{\overline{s}_{\alpha+1}} \right), \nonumber
\end{eqnarray} 
where the reduced density matrix and the auxiliary operators enter as $\sigma^{(0)}(t)=\sigma(t)$ and $\sigma^{(1)}_{j_{1}}(t)=\sigma_{K,n,s,p}(t)$, 
using superindices $j_{\beta}=(K_{\beta},m_{\beta},s_{\beta},p_{\beta})$ for notational reasons ($\alpha,\beta\in\mathbb{N}$). 
The higher tier operators $\sigma^{(\alpha)}_{j_{1}..j_{\alpha}}(t)$ ($\alpha\geq2$) are associated with the nested commutators 
$ \left[ H_{\text{tun}}(t) , \left[ H_{\text{tun}}(t) , .... , \varrho(t) \right] \right]$.  
They can be represented by a set of superoperators, which are defined by 
\begin{eqnarray}
B_{j} \varrho(t) &\equiv& \sigma^{(1)}_{j_{1}}(t), 
\end{eqnarray}
as 
\begin{eqnarray}
\label{defauxop} 
 \sigma^{(\alpha+1)}_{j_{1}..j_{\alpha+1}}(t) &=&  B_{j_{\alpha}} .. B_{j_{1}} \varrho(t) .  
\end{eqnarray}
They enter the hierarchy (\ref{hierarcheom}) with prefactors $\nu_{K_{\alpha+1},m_{\alpha+1}} \nu_{K_{\alpha+1},n_{\alpha+1}}/\gamma$ 
such that a  truncation of the hierarchy at the $\alpha$th tier leads to an equation of motion for the reduced density matrix  
that is valid up to the $\alpha$th order in the hybridization strength $\varGamma=\text{max}(\nu_{K,m} \nu_{K,n}/\gamma)$. 
The corresponding dimensionless expansion parameter is the ratio of $\varGamma$ to the minimal value of $\text{Re}\left[\omega^{\pm}_{K,p}\right]$, 
which is $\sim\text{min}(\gamma, k_{\text{B}}T)$. Thus, for example, the approach can be expected to describe Kondo physics 
if the ratio  $\varGamma/ (k_{\text{B}}T_{\text{Kondo}})$ of the hybridization strength and the 
Kondo temperature is smaller or comparable to one (although convergence may be achieved for larger values as well) \cite{Zheng2009,Yan2012}. 
Further details on how the hierarchy of equations of motion (\ref{hierarcheom}) is solved numerically are given in App.\ \ref{appB}.

While our approach closely follows that of Jin \emph{et al.}\ \cite{Jin2008} we  redefined the auxiliary operators (cf.\ Eq.\ \ref{defauxop2})  
in a dimensionless way and formulated an improved measure for truncating the hierarchy of  equations of motion 
systematically (see App.\ \ref{appB}). Thereby, in contrast to earlier work \cite{Shi2009}, our methodology respects the internal 
structure of both the auxiliary operators and the corresponding equations of motion. Moreover, our choice of the level-width 
functions (cf.\ Eq.\ \ref{levelwidths}) means that, by working in a basis where 
$\Gamma_{\text{L},mn}(\epsilon)\sim \delta_{m,1}\delta_{n,1}$ and 
$\Gamma_{\text{R},mn}(\epsilon)=0$ for $m,n>2$, 
the state variables $m_{\beta}$ and $n_{\beta}$ become redundant. In the present context, this reduces the number of superindices 
from $2\cdot N_{\text{leads}}\cdot N_{\text{el}}^{2}\cdot N_{p}$ to  $2\cdot (N_{\text{leads}}+1)\cdot N_{p}$, 
where $N_{\text{leads}}$, $N_{\text{el}}$ and $N_{p}$ denote the number of  
electrodes, eigenstates of the system $S$ and the number of terms considered in the decomposition (\ref{paramscheme}), 
respectively\footnote{In the systems considered in this article, $\Gamma_{\text{L},mn}(\epsilon)\sim \delta_{m,1}\delta_{n,1}$ and 
$\Gamma_{\text{R},mn}(\epsilon)\sim \delta_{m,2}\delta_{n,2}$ holds  such that the number of superindices reduces even 
further to $2\cdot N_{\text{leads}}\cdot N_{p}$.}.

In addition to the Matsubara decomposition (\ref{paramscheme}), other schemes are based 
on, for example, Gauss-Legendre or Gauss-Chebyshev quadrature schemes of the 
Fourier transforms of $C^{s}_{K,mn}(t)$ \cite{Heng2012}, or on 
hybrid schemes of the latter and the Matsubara decomposition scheme \cite{Zheng2009} or on
Pade approximation schemes \cite{Ozaki2007,Hu2010,Hu2011}. Thus, the efficiency of the HQME method can  
be somewhat increased (see the discussion at the end of Sec.\ \ref{numerisec}). However, the basic methodological 
characteristics, in particular that the method is in effect an expansion in $\varGamma/\text{min}(\gamma, k_{\text{B}}T)$, are 
maintained. Further theoretical progress is required to address the moderate to strong coupling regime,  
where the hybridization is comparable to or larger than the temperature scales of the system (such as, \emph{e.g.}, deep 
inside the Kondo regime).

\subsection{Born-Markov master equation approach} 
\label{bornmarkovtheory}

The long-time limit of the reduced density matrix $\sigma$ is often approximated as the stationary solution of the 
well-established equation of motion \cite{May02,Mitra04,Lehmann04,Harbola2006,Volkovich2008,Hartle09,Hartle2010,Hartle2010b}
\begin{eqnarray}
\label{genfinalME}
\frac{\partial \sigma(t)}{\partial t} &=& -i \left[ 
H_{\text{S}} , \sigma(t) \right]  - \int_{0}^{\infty} \text{d}\tau\, 
\text{tr}_{\text{L+R}}\lbrace \left[ H_{\text{tun}} , \left[ \tilde{H}_{\text{tun}}(\tau), \sigma(t) 
\sigma_{\text{L+R}} \right] \right] \rbrace , 
\end{eqnarray}
with 
\begin{eqnarray}
\tilde{H}_{\text{tun}}(\tau) = 
\text{e}^{-i(H_{\text{S}}+H_{\text{L}}+H_{\text{R}})\tau} 
H_{\text{tun}} \text{e}^{i(H_{\text{S}}+H_{\text{L}}+H_{\text{R}})\tau}.
\end{eqnarray}
Here, $\sigma_{\text{L+R}}$ represents the equilibrium density matrix of the leads. Eq.\ (\ref{genfinalME}) can be 
derived from the Nakajima-Zwanzig equation \cite{Nakajima,Zwanzig}, 
employing a second-order expansion in the coupling $H_{\text{tun}}$ along with the so-called Markov approximation. 
Note that the master equation (\ref{genfinalME}) corresponds to the Redfield (or Bloch-Wangsness-Redfield) 
equation if it is evaluated in the eigenbasis of the system Hamiltonian $H_\text{S}$ \cite{Wangsness1953,Redfield1965,May04}.

The BM master equation (\ref{genfinalME}) is commonly evaluated by neglecting the real part $\mathbf{\Delta}$ of the 
self-energy matrix $\mathbf{\Sigma}=\mathbf{\Delta}-(i/2)\mathbf{\Gamma}$, 
which is associated with the coupling between the system S and the electrodes L and R. Thus, the renormalization of 
the energy levels 1 and 2 due to the coupling to  the leads (given by the diagonal elements $\Delta_{11}$ and $\Delta_{22}$  
\cite{Harbola2006,Hartle2010b}) is neglected. A coupling between these states mediated by the off-diagonal elements $\Delta_{12}$ 
is also neglected. For systems without quasidegenerate levels, these terms are not important \cite{Hartle2010b} 
but in the present situation we will see by comparison to the HQME method (vide infra) that this inter-state coupling 
can play an important role for the transport properties.

\subsection{Nonequilibrium Green's function approach based on the non-crossing approximation}
\label{ncatheory}

The NCA is a popular and successful method for calculating the nonequilibrium Green's 
functions of quantum dots\cite{Keiter1970,Keiter1971,Grewe1981,Kuramoto1983,Coleman1984,Bickers1987b,Bickers1987,Nordlander1999,Eckstein2010,Gull2011}.  
The NCA involves a hybridization expansion, where non-crossing diagrams of all orders are summed in terms of a Dyson 
series, but diagrams with crossing hybridization lines are excluded. As an approximate method, it should always be 
considered less reliable than converged numerically exact results like HQME and quantum Monte Carlo schemes 
\cite{Gull2011,Eckstein2010,Cohen2013b}. Nevertheless, the NCA can capture qualitative aspects of the physics involved 
in this problem and represents a standard methodology for the description of
electron-electron interaction effects. It is thus very useful to compare exact results obtained with HQME to the ones 
obtained with NCA. In the present context, the simplicity and flexibility of the NCA even 
facilitates a study of the effects for different lead density of states (see Sec.\ \ref{ncaresults}).

\begin{figure}
\includegraphics[width=11cm]{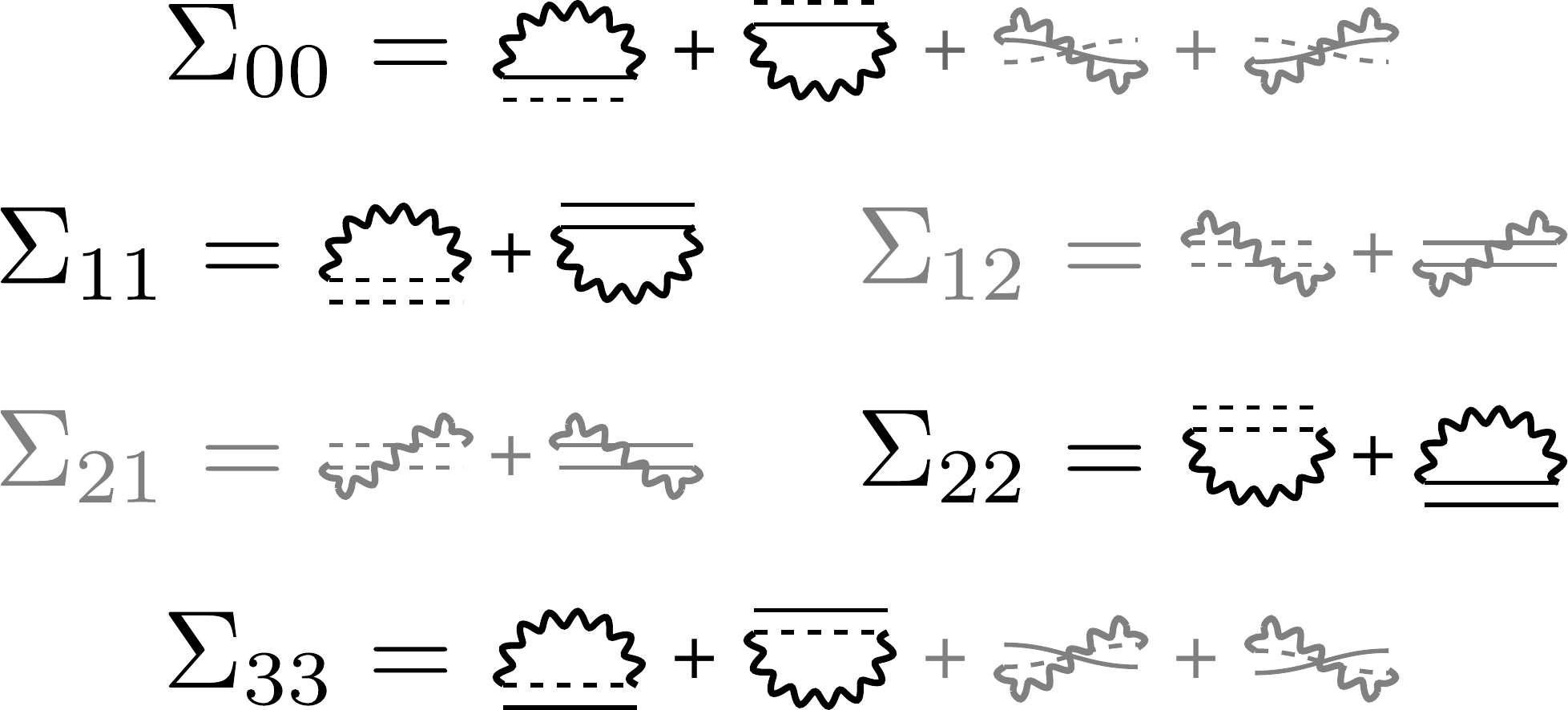}\caption{NCA self-energy diagrams for the spinless Anderson model; gray terms
are identically zero in the regular Anderson model.\label{NCA_diagrams}}
\end{figure}

In order to extend the NCA framework to the spinless Anderson model, we consider  the Dyson equation for the causal 
propagator $G\left(t\right)=\mathrm{Tr}_{\text{L+R}}\left\{ \rho\left(t=0\right)e^{-iHt}\right\} $:
\begin{equation}
\label{ncaprop}
G\left(t-t^{\prime}\right)=G^{\left(0\right)}\left(t-t^{\prime}\right) + 
\int_{t^{\prime}}^{t}\mathrm{d}t_{1}\int_{t^{\prime}}^{t1}\mathrm{d}t_{2}G^{\left(0\right)}\left(t-t_{1}\right)\Sigma\left(t_{1},t_{2}\right)G\left(t_{2}-t^{\prime}\right).
\end{equation}
Here, $G^{\left(0\right)}$ is the  propagator for the case where the dot-lead coupling is set to zero and the self-energy 
$\Sigma$ induces all diagrams arising from the coupling to the electrodes.  The matrix elements 
$G_{ab}\equiv\left\langle a\right|G\left|b\right\rangle $ of the propagator in the basis of the dot states can be 
conveniently represented by a pair of lines (one for the occupation of each spin
level, cf.\ Fig.\ \ref{NCA_diagrams}), where a dashed line stands for an empty level  and a solid line for a full one.

In the  spinful Anderson model, $G$ is diagonal in this basis and pairs of interactions with the electrodes can be 
represented diagrammatically as wavy lines which flip one spin between interactions (when using the shorthand diagrams 
in which the many-body state on the dot is represented by a single line, the sum over the two arcs is often compactly 
represented by a single one). The definition of the NCA self-energy is then given by the black diagrams in Fig.~\ref{NCA_diagrams}.

The new feature of the model we consider is that  non-diagonal ($m\neq n$) contributions to $\Gamma_{K,mn}(\epsilon)$ 
are also present. These induce a new kind of hybridization line, which takes an electron from  one level to the other. 
The additional diagrams are shown in gray in Fig.\ \ref{NCA_diagrams};  note that, as a result, 
$G_{ab}$ for $\left|a\right\rangle =\left|01\right\rangle ,\,\left|b\right\rangle =\left|10\right\rangle $ and vice-versa 
become non-zero and are represented by pairs of lines in which the level occupations trade places. The need to store and 
compute additional matrix-elements in the spinless case complicates the calculation somewhat, but no conceptual differences 
arise. Similar NCA treatments have been used to study multi-orbital models \cite{Lorente2011,Korytar2012} and the spinless 
Anderson model \cite{RouraBas2010,RouraBas2011,Trocha2012}. However, it is important to note that 
what the authors of these works refer to as the NCA is an infinite-$U$ approach which differs from the 
finite-$U$ method employed here.

\subsection{Observables of interest}

The crucial quantities that characterize the nonequilibrium transport properties of interest here are the steady state level  
populations $n_{1/2}$, the inter-level  coherence $\sigma_{10,01}$ and the electrical current that is flowing through the system S. 
In the occupation number representation,  the population 
of the electronic levels is  given by the diagonal elements of the reduced density matrix  
\begin{eqnarray}
 n_{1} &=& \lim_{t\rightarrow\infty} \left( \sigma_{10,10}(t) + \sigma_{11,11}(t) \right), \\
 n_{2} &=& \lim_{t\rightarrow\infty} \left( \sigma_{01,01}(t) + \sigma_{11,11}(t) \right),   
\end{eqnarray}
while the off-diagonal elements of the reduced density matrix determine the coherence 
\begin{eqnarray}
 \sigma_{10,01} &=& \lim_{t\rightarrow\infty}  \sigma_{10,01}(t).  
\end{eqnarray}
The coherence has no classical analog; it encodes quantum mechanical tunneling and interference effects.  For well 
separated energy levels, coherences of the density matrix vanish.  For degenerate or quasidegenerate levels, however, 
coherences can become as important as the populations $n_{1/2}$. In the present context, the inter-level coherence 
can be used as a measure for the strength of interference effects, that is comparison of the coherence found in the interacting 
and non-interacting situations provides an assessment of the importance of interaction-induced decoherence.

The electrical current flowing through the systems S is determined by the number of electrons that enter or leave the   
electrode $K$ in a given interval of time ($K\in\lbrace \text{L,R}\rbrace$) 
\begin{eqnarray}
\label{firstcurrent}
I_{K} &=& - e \frac{\text{d}}{\text{d} t} 
\sum_{k\in K} \langle c^{\dagger}_{k} c_{k} \rangle \equiv \langle \hat{I}_{K} \rangle, 
\end{eqnarray}
where $-e$ denotes the charge of an electron.  It can be written in terms of the first-tier auxiliary operators $\sigma^{(1)}_{j}(t)$ as \cite{Jin2008} 
\begin{eqnarray}
 I_{K} &=& e  \sum_{K,m,n,p}  \frac{\nu_{K,m} \nu_{K,n}}{\gamma} \lim_{t\rightarrow\infty} \left( \text{Tr}_{\text{S}}\left[  \sigma_{K,m,+,p}(t) d_{n} \right] 
- \text{Tr}_{\text{S}}\left[ d_{n}^{\dagger} \sigma_{K,m,-,p}(t) \right] 
\right). 
\end{eqnarray}
or, using the BM master equation scheme (see Sec.\ \ref{bornmarkovtheory}), as 
\begin{eqnarray}
\label{gencurrentME}
I_{K} &=& -i \lim_{t\rightarrow\infty} \int_{0}^{\infty}\text{d}\tau\, \text{tr}\lbrace 
\left[ \tilde{H}_{\text{tun}}(\tau) , \sigma_{\text{L+R}}  \sigma(t) \right] \hat{I}_{K} \rbrace .
\end{eqnarray}

\begin{figure}
\includegraphics[width=13cm]{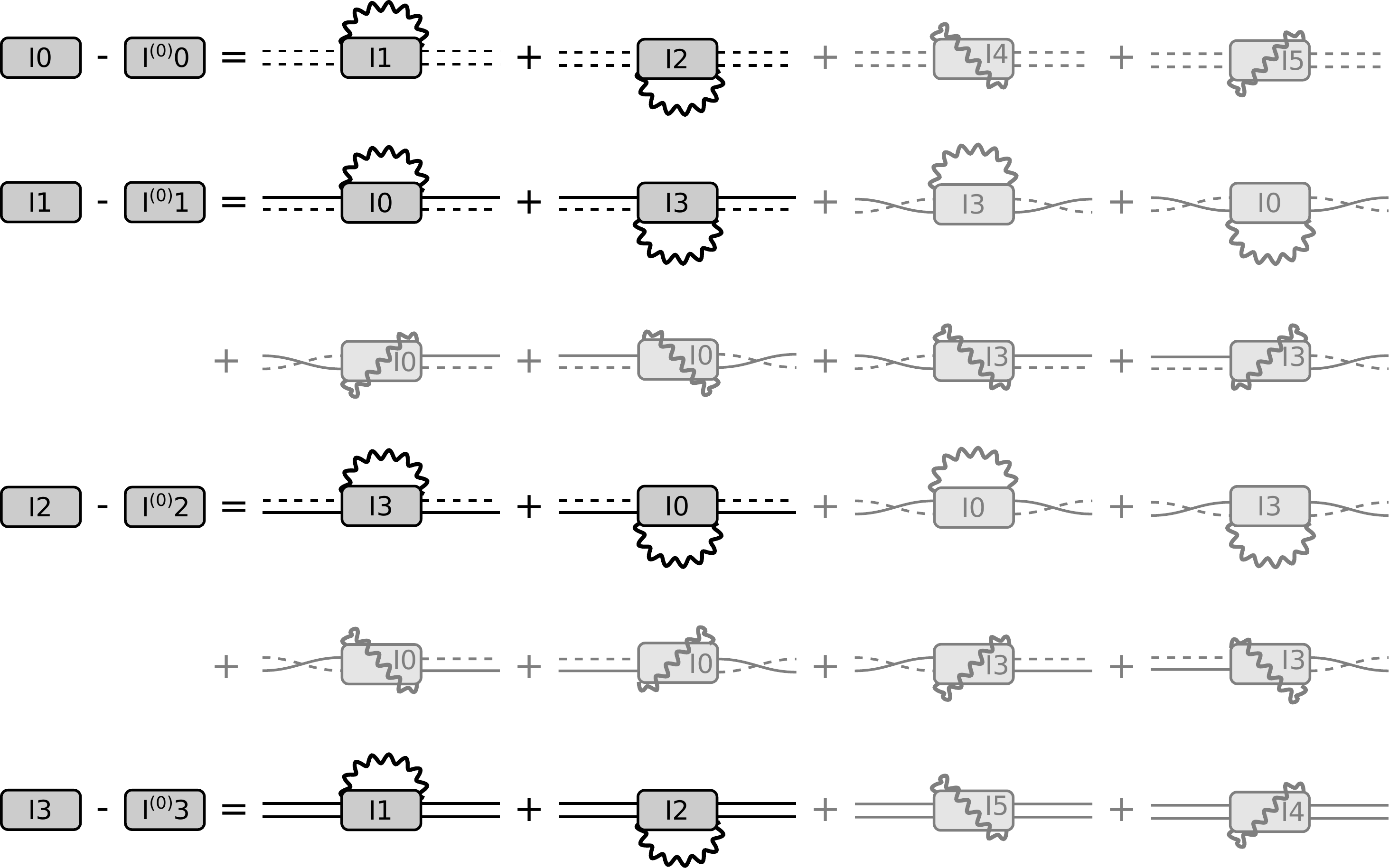}\caption{NCA diagonal vertex diagrams for the spinless Anderson model; 
gray
terms are identically zero in the regular Anderson model.\label{NCA_vertices_diagonal}}
\end{figure}

\begin{figure}
\includegraphics[width=13cm]{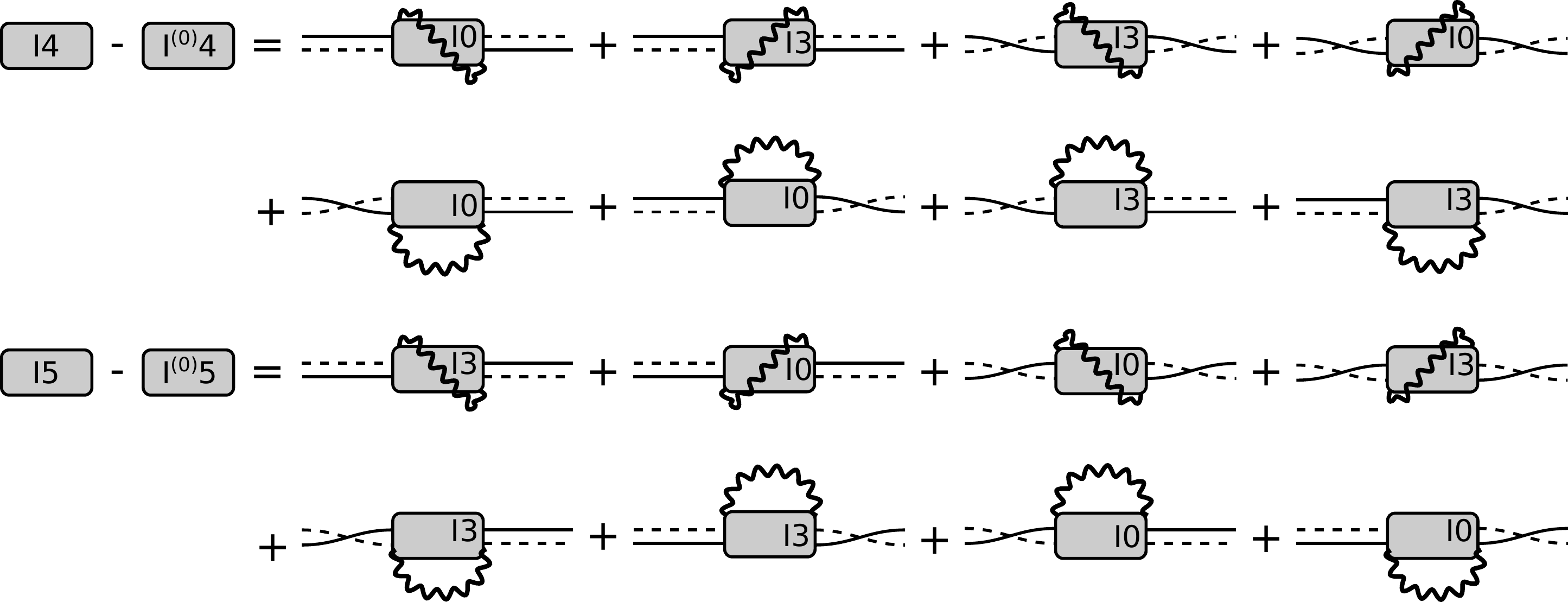}\caption{NCA off-diagonal vertex diagrams for the spinless Anderson 
model;
all the terms appearing here are identically zero in the regular Anderson
model.\label{NCA_vertices_off_diagonal}}
\end{figure}

The computation of the electronic state populations, the coherence and the electrical current within the NCA framework 
requires the solution of a set of secondary vertex equations. A vertex is defined in this context as an object which 
exists on both parts of the contour simultaneously (cf.\ Figs.\ \ref{NCA_vertices_diagonal} and \ref{NCA_vertices_off_diagonal}).  
It is specified by an inner vertex, which is associated with the corresponding observable  (for instance, the inner vertex 
corresponding to measuring the population  in the state $\left|a\right\rangle $ is given by  
$P_{a b}\left(t,t^{\prime}\right)=G_{aa}\left(t\right)G_{aa}^{\dagger}\left(t^{\prime}\right)\delta_{a b}$) and outer indices, 
which connect the inner vertex to the initial state of the system via the propagator (\ref{ncaprop}). 
Together with the specification of the observable, the vertex equations form a set of coupled integral equations which when 
solved together  provide a self-consistent resummation of all contour-crossing NCA diagrams. The vertex terms which appear 
in the original Anderson case are the dark diagrams in Fig.~\ref{NCA_vertices_diagonal}; off-diagonal vertex contributions 
are unique to the spinless case and shown in Fig.\ \ref{NCA_vertices_off_diagonal}. An alternative scheme, formulated in 
terms of correlation functions instead of propagators, has been proposed  \cite{Eckstein2010} but will not be used here.

\section{Results}
\label{results}

We study two complementary realizations of the spinless Anderson model, which we refer to as DES and CON. The DES realization 
may represent a linear (molecular) conductor \cite{Solomon2008b,Hartle2012} or  two quantum dots connected in series. 
The second system, model CON, may represent a branched (molecular) conductor \cite{Brisker2008,Hartle2012} or two quantum dots connected in parallel 
\cite{Holleitner2001,Holleitner2002,Iye2001,Gefen2002,Kubala2002,Cohen2007,Wegscheider2007,Segal2012}. 
Mathematically the two models are distinguished by the coupling parameters, which 
reflect the different connectivities of the eigenstates and the leads. In model DES these coupling parameters are  
symmetric, $\nu_{\text{L},1}=\nu_{\text{R},1}$, and antisymmetric, $\nu_{\text{L},2}=-\nu_{\text{R},2}$, respectively 
(the corresponding dot-lead coupling parameters (cf.\ Fig.\ \ref{LinConduct}) are $\nu_{a/d}=\sqrt{\nu_{\text{L/R},1}^{2}+\nu_{\text{L/R},2}^{2}}$ and $\nu_{b/c}=0$). 
On the single-particle level it 
has been shown that this form of the coupling produces strong destructive interference effects which suppress the 
current flow \cite{Hartle2011b,Hartle2012}. In model CON the coupling parameters $\nu_{K,1/2}$ are the same or, equivalently, symmetric 
($\nu_{a/b}=\sqrt{\nu_{\text{L/R},1}^{2}+\nu_{\text{L/R},2}^{2}}$ and $\nu_{c/d}=0$). 
Thus, the system shows constructive 
interference effects \footnote{Note that a clear separation between systems that show constructive and destructive 
interference effects  is, in general, not possible \cite{Hartle2012}.}.  A detailed list of model parameters is given 
in Tab.\ \ref{parameters}. Note that these model parameters reflect typical experimental values 
\cite{Holleitner2001,Wegscheider2007,Osorio2007,Nilsson2010} 
with respect to the temperature scale $k_{\text{B}}T\approx25$\,meV that is used in this article.  

\begin{table}
\begin{center}
\begin{tabular}{|*{11}{ccc|}}
\hline \hline 
&model &&& $\epsilon_{1}$&&&$\epsilon_{2}$&&&$U$&&& $\nu_{\text{L},1}$&&& $\nu_{\text{L},2}$&&& $\nu_{\text{R},1}$&&&$\nu_{\text{R},2}$&&&
$\gamma$&\\ \hline 
& DES &&&0.5&&&0.501&&&0.5&&&$\nu$&&&$\nu$&&&$-\nu$&&&$\nu$&&&2&\\
& CON &&&0.5&&&0.501&&&0.5&&&$\nu$&&&$\nu$&&&$\nu$&&&$\nu$&&&2&\\
& CENTRAL &&&0.5&&&0.75&&&0.5&&&$\nu$&&&$\nu$&&&$\nu/3$&&&$\nu/3$&&&2&\\
& BLOCK &&&0.5&&&0.75&&&0.5&&&$\nu$&&&$\nu$&&&$\nu$&&&$\nu/3$&&&2&\\
\hline \hline
\end{tabular}
\end{center}
\caption{\label{parameters}
Parameters for the double dot devices that are investigated in this article. All energy values are given in $\mathrm{eV}$, 
the temperature of the electrodes $T$ is set to $300$\,K ($\simeq25$\,meV) and the value of the dot-lead coupling 
parameter is $\nu = 42$\,meV.  
}
\end{table}

\subsection{Transport properties of junction DES}
\label{dessec}

\begin{figure}
\begin{tabular}{l}
\hspace{-0.5cm}(a) \\
\resizebox{\newwidth}{\newheight}{
\includegraphics{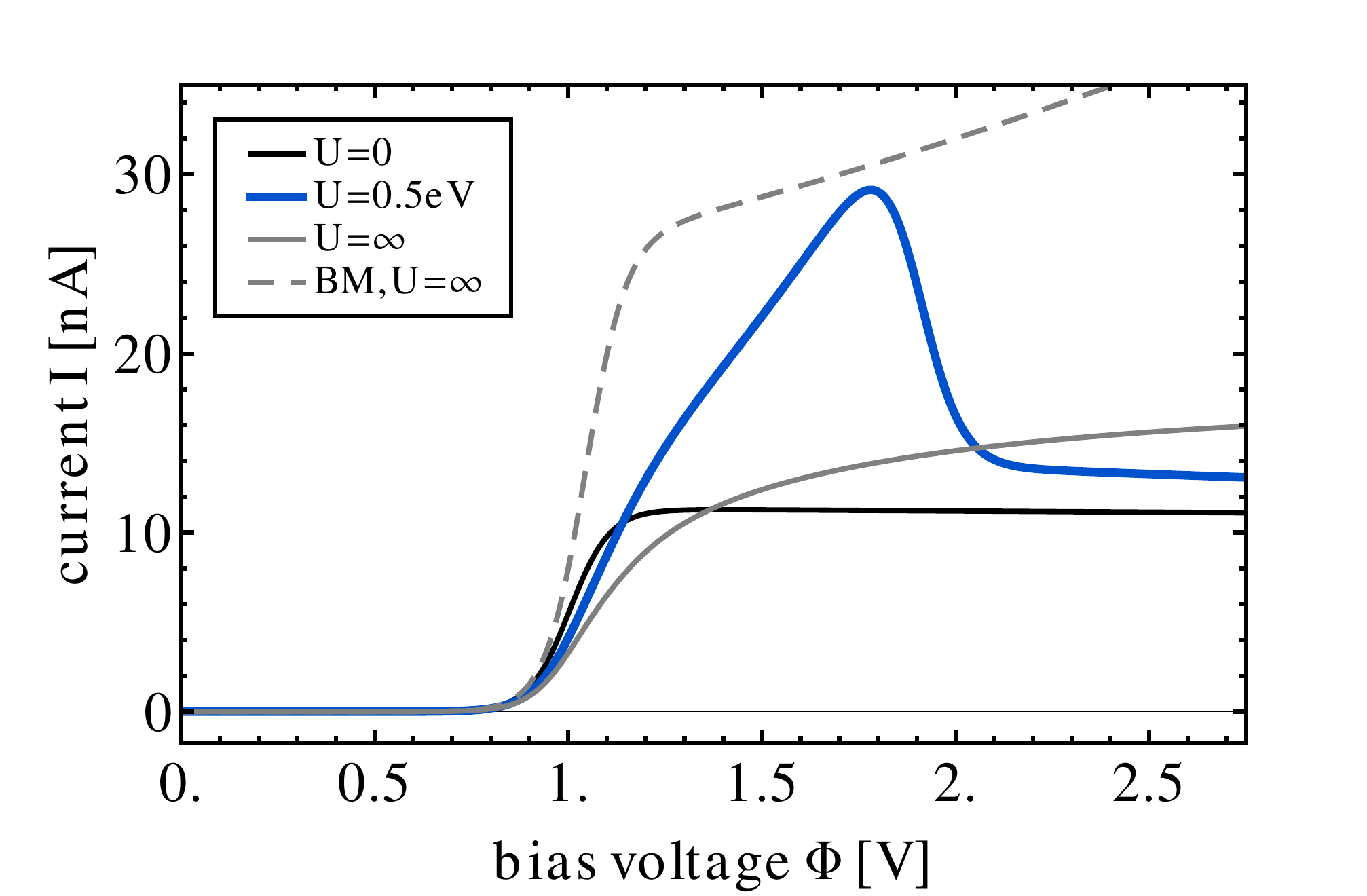} 
}\\
\hspace{-0.5cm}(b) \\
\resizebox{\newwidth}{\newheight}{
\includegraphics{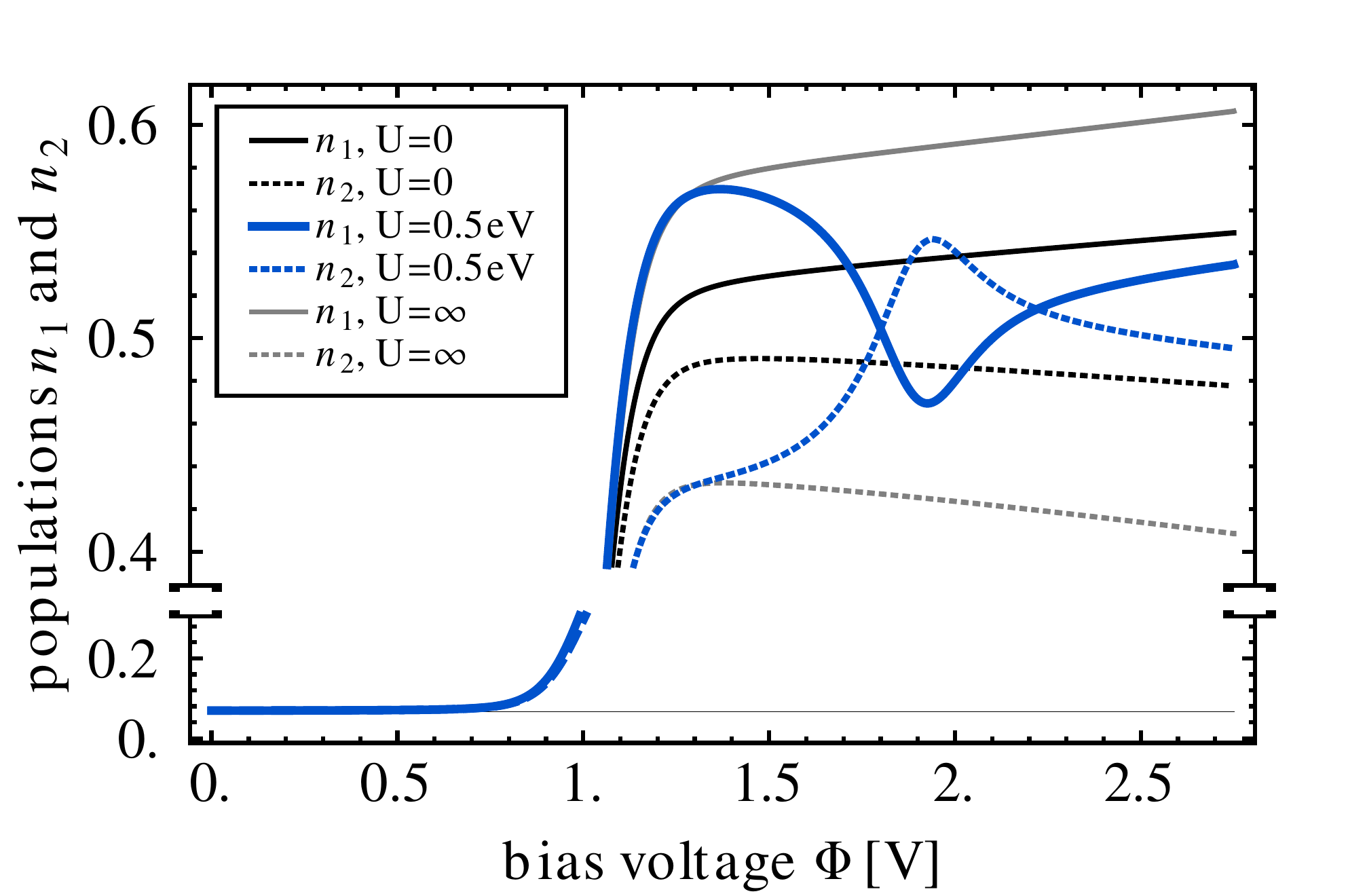}
}
\end{tabular}
\caption{(Color online)\label{model6current} Current-voltage and electronic population characteristics of junction DES calculated for 
different electron-electron interaction strengths: $U=0$ (black lines), $U=\infty$ (gray lines) and 
$U\approx20 k_{\text{B}}T$ (blue lines). The solid lines have been  obtained using the HQME method. The dashed 
line has been computed with the BM scheme. Negative differential resistance occurs in the same voltage range where an   
inversion of the electronic populations $n_{1}$ (solid lines) and $n_{2}$ (dotted lines) occurs. 
}
\end{figure}

The calculated current-voltage characteristic of junction DES is shown as the solid blue line in Fig.\ \ref{model6current}a.   
The current increases monotonically with bias voltage $\Phi$ for $\Phi\lesssim1.7$\,V. The monotonic increase occurs both in the 
non-resonant transport regime ($0<e\Phi\lesssim2\epsilon_{1/2}\approx1$\,eV), where both eigenstates of the junction are located 
outside the bias window, \emph{i.e.}\ $\epsilon_{1/2}>\mu_{K}$, and in part of the  resonant transport regime 
($2\epsilon_{1/2}\lesssim e\Phi \lesssim 2\epsilon_{1/2}+U\approx1.5$\,eV) where the two levels are located within the bias window, 
\emph{i.e.}\ $\mu_{\text{L}}>\epsilon_{1/2}>\mu_{\text{R}}$. As the voltage is increased beyond $\Phi\approx1.7$\,V, junction 
DES shows a voltage range with a pronounced decrease of the electrical current as voltage is increased, in other words a negative 
differential resistance which for the parameters considered here is peaked at $\Phi\approx1.9$\,V.

The corresponding populations of the electronic levels $n_{1}$ and $n_{2}$ are shown  as the solid and dotted blue 
lines in Fig.\ \ref{model6current}b, respectively. The two levels are almost unpopulated in the non-resonant 
transport regime. In the resonant transport regime, the chemical potential in one of the leads  exceeds the level 
positions and, accordingly, the population of the two levels increases substantially such that the average number of electrons 
in the junction is close to $n_{1}+n_{2}\approx1$. 
Despite the near degeneracy of the two levels, their populations  differ significantly. Moreover, an inversion of 
the population occurs for the  bias voltages where the decrease of the current level  is most pronounced.

We now show  that this behavior is associated with the quenching of destructive interference  effects due to 
electron-electron interactions and the energy-dependence of the density of states in the electrodes.  
The argument has three steps. First, we study the case where no electron-electron interactions are present 
in the system ($U=0$) and destructive interference  effects are fully developed. Second, we consider the 
limit $U\rightarrow\infty$, where destructive interference effects appear  to be quenched. In the third and 
final step, we  consider a finite value for $U$ and show that the quenching of interference effects  due to 
electron-electron interactions is no longer effective once the bias voltage exceeds $\approx2(\epsilon_{1/2}+U)$.

\subsubsection{The non-interacting limit}
\label{desnonint}

\begin{figure}
\resizebox{\newwidth}{\newheight}{
\includegraphics{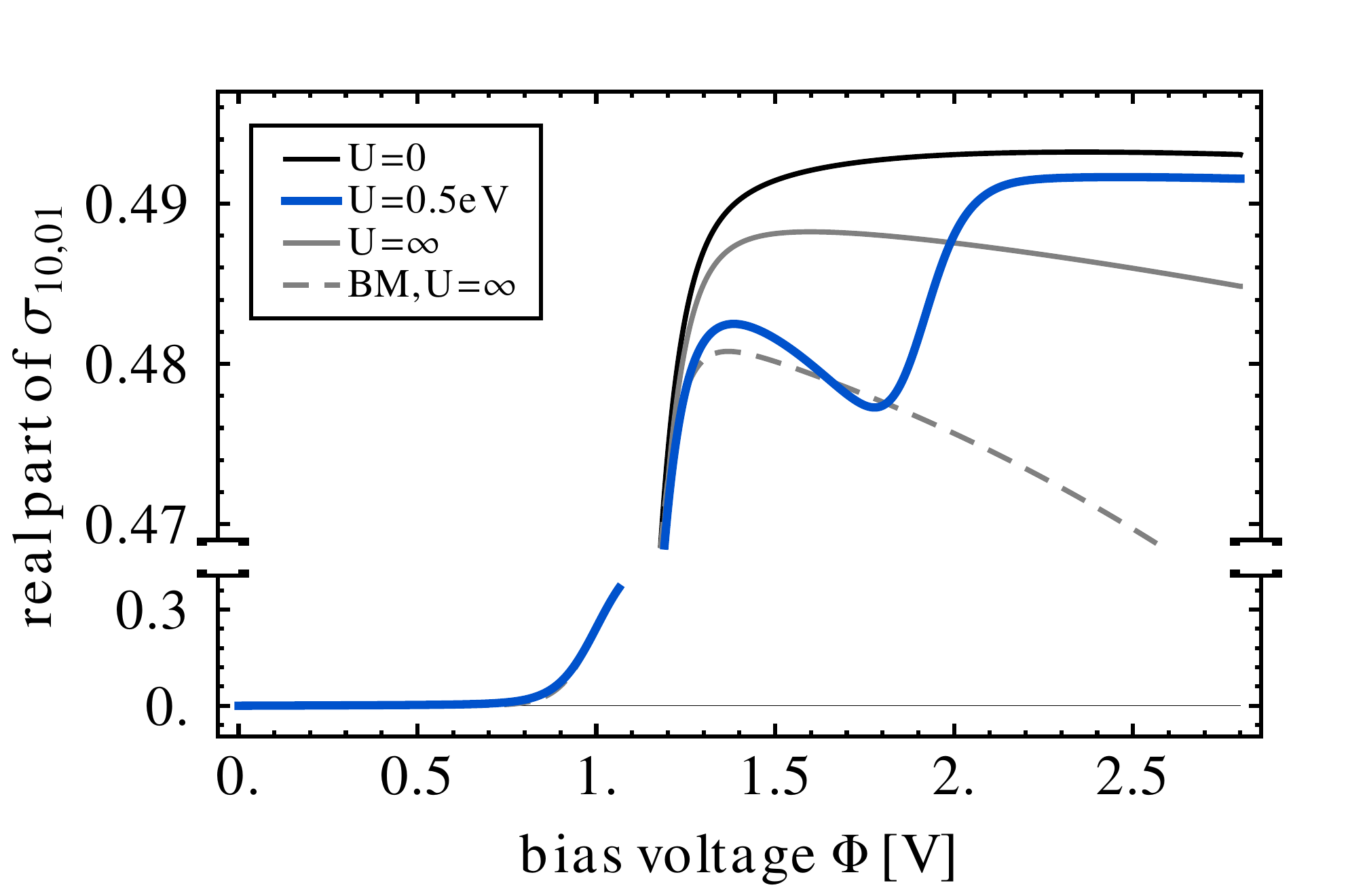}
}
\caption{(Color online)\label{model6currentX} Real part of the coherence $\sigma_{10,01}$ corresponding to the 
characteristics shown in Fig.\ \ref{model6current}. Comparison to the non-interacting case (black line)   
quantifies the effect of decoherence due to electron-electron interactions. 
}
\end{figure}

If electron-electron interactions are neglected in model DES, one obtains the current-voltage characteristic 
represented by the solid black line in Fig.\ \ref{model6current}a. It shows a single step at 
$e\Phi\approx2\epsilon_{1/2}$ that indicates the onset of resonant transport through states 1 and 2. 
The low value of the current is due to destructive interference effects \cite{Solomon2008b,Hartle2012}. 
This can be inferred from  an analysis in terms of a BM master equation scheme  (cf.\ Sec.\ \ref{bornmarkovtheory}) 
\cite{May02,Mitra04,Harbola2006,Hartle2010b}.  Using Eqs.\ A1 and A4 of Ref.\ \cite{Hartle2010b}, the current through 
junction DES can be written as 
\begin{eqnarray}
\label{bornmarkovcurrent}
 I &=& 2 e \Gamma \left( \sigma_{10,10} + \sigma_{10,10} - 2\text{Re}\left[\sigma_{10,01}\right] \right),   
\end{eqnarray}
where $\Phi>2\epsilon_{1/2}$ is assumed and the wide-band approximation with $\Gamma=\Gamma_{\text{L},11}(\mu_{\text{L}})$ 
is employed. The first two terms are given by the population of the singly occupied states  $\sigma_{10,10}$ and 
$\sigma_{10,10}$, respectively. They represent the incoherent sum current that is flowing through the two states 
and are approximately given by the sum of the populations $n_{1}+n_{2}\approx \sigma_{10,10} + \sigma_{10,10}$. 
In the resonant transport regime, this sum is $\approx1$, indicating that the bottleneck for transport is given by 
the small inter-dot coupling $(\epsilon_{2}-\epsilon_{1})/2$ rather than by the coupling to the electrodes 
(\emph{i.e.}\ the system S is mostly occupied by a single electron). The last term is given by the real part of the 
coherence $\sigma_{10,01}$. It encodes the effect of destructive interference  and is depicted in Fig.\ \ref{model6currentX}. 
As can be seen, it almost cancels the first two terms, since its value is very close to $0.5$ in the resonant 
transport regime. The real part of the coherence $\sigma_{10,01}$ can thus be used as a measure for the strength 
of destructive interference.

While the suppression of the current flow in systems like junction DES is well known \cite{Solomon2008b,Hartle2012},  
it is less recognized that in this model the population of the electronic levels is significantly different. 
The two level occupancies  are depicted in Fig.\ \ref{model6current}b by the solid black and the dotted black line, 
respectively. The voltage dependence of the level occupancies is in general similar to that of the corresponding 
current-voltage characteristic, in particular exhibiting a step at $e\Phi\approx2\epsilon_{1/2}$. 
Despite the near degeneracy of the level energies, the corresponding populations are not the same, where, 
for voltages larger than $e\Phi\approx2\epsilon_{1/2}$, the population difference 
even increases with the applied bias voltage.

An analysis of the retarded/advanced single-particle Green's functions $G^{\text{r/a}}_{mn}(\epsilon)$ reveals 
the origin of this behavior. They are defined by \cite{Haug98}
\begin{eqnarray}
 \textbf{G}^{\text{r/a},-1}_{mn}(\epsilon) &=& (\epsilon- \epsilon_{m}) \delta_{mn} 
- \Delta_{K,mn}(\epsilon) + \frac{i}{2} \Gamma_{K,mn}(\epsilon). 
\end{eqnarray}
Thereby, the functions $\Delta_{K,mm}(\epsilon)$ denote the renormalization of the energy levels $\epsilon_{m}$ due to 
the coupling of the junction to lead $K$ and the off-diagonal elements  $\Delta_{K,mn}(\epsilon)$ ($m\neq n$) encode a inter-state 
coupling which is induced by the energy dependence of 
$\Gamma_{K,12}(\epsilon)$. In the present context, the energy dependence is due to the finite bandwidth $\gamma$. 
In general, it may also be the result of an energy dependence of the couplings $\nu_{K,m}$. 
Using the Green's functions $\textbf{G}^{\text{r/a}}_{mn}(\epsilon)$, the population of 
the two levels can be calculated according to the formula 
\begin{eqnarray}
 n_{m} &=& \int\frac{\text{d}\epsilon}{2\pi} \sum_{Kno} G^{\text{r}}_{mn}(\epsilon) \Gamma_{K,no}(\epsilon) G^{\text{a}}_{om}(\epsilon) f_{K}(\epsilon).  
\end{eqnarray}
This procedure yields exactly the same result as the one that is obtained by the 
HQME method outlined in Sec.\ \ref{heomtheory}. However, if the off-diagonal elements $\Delta_{K,12}(\epsilon)$ 
are neglected  
\footnote{Note that this cannot be done in a systematic way within the HQME approach, as the $\Delta_{K,mn}(\epsilon)$ 
do not enter this formalism explicitly}, 
the populations of the two levels become very similar ($n_{1}-n_{2}<10^{-3}$ for $\Phi>2\epsilon_{12}+k_{\text{B}}T$). 
This shows that the off-diagonal elements $\Delta_{K,12}(\epsilon)$ 
cause the pronounced asymmetry in the electronic populations $n_{1}$ and $n_{2}$  
\footnote{In addition, we have crosschecked these findings with respect to other kinds of spectral functions, 
such as, for example, a semi-elliptic conduction band, and obtained very similar results. 
The same holds true if $\Delta_{K,12}(\epsilon)$ is replaced by a constant.}. 
Note that Eq.\ (\ref{bornmarkovcurrent}) is derived from Born-Markov theory, where the effect 
of the off-diagonal elements $\Delta_{K,12}(\epsilon)$ is neglected such that the corresponding 
difference in the electronic populations $n_{1}$ and $n_{2}$ is small, that is the 
relation $I>2e\Gamma \sqrt{\vert n_{1}-n_{2}\vert}$ (which can be derived using the 
relation $\vert \rho_{10,01} \vert<\sqrt{\rho_{10,10}\rho_{01,01}}$) is fullfilled.

\subsubsection{Decoherence phenomena in the limit $U\rightarrow\infty$}

\label{uinfsec}

Now, we consider junction DES in the limit $U\rightarrow\infty$. The gray lines in Fig.\ \ref{model6current}b show   
the corresponding electronic populations $n_{1}$ (solid) and $n_{2}$ (dotted). The difference in the electronic 
populations is seen to be considerably larger at $U=\infty$ than it is at $U=0$.  The $U$-dependence arises because,  
at $U=\infty$, transport through one of the two levels is completely blocked whenever the other level is occupied.

The current-voltage characteristic at $U=\infty$  is shown as the solid gray line in Fig.\ \ref{model6current}a. 
At $U=\infty$ the current  is smaller than the $U=0$ current at low and intermediate bias voltages, \emph{i.e.}\ for $\Phi<1.3$\,V, but 
is larger for higher bias voltages and continues to increase as $\Phi$ is increased further into the resonant transport 
regime. The increase of the current is accompanied by a decrease of the real part of the 
coherence $\sigma_{10,01}$ 
(compare the solid gray and black lines in Fig.\ \ref{model6currentX}), indicating that the quenching of destructive interference effects, 
which are suppressing the current flow in this system, becomes progressively stronger as the voltage is increased.

This decoherence effect can be qualitatively understood via a Born-Markov analysis similar to that previously given for 
the non-interacting case  (cf.\ Eq.\ \ref{bornmarkovcurrent}). 
In the limit $U\rightarrow\infty$ and in the Born-Markov approximation, the current is given by 
\begin{eqnarray}
\label{bornmarkovcurrent2}
 I &=& e \Gamma \left( \sigma_{10,10} + \sigma_{10,10} - 2\text{Re}\left[\sigma_{10,01}\right] \right)  . 
\end{eqnarray}
Comparison to Eq.\ (\ref{bornmarkovcurrent}) shows that within the BM approximation the $U=\infty$ current is a factor 
of two smaller than the $U=0$ current. However, a decrease in the real part of the coherence $\sigma_{10,01}$ overrides 
this effect, so that the net result is an enhancement of the current (cf.\ Figs.\ \ref{model6current}a and \ref{model6currentX}). 
The difference with respect to the non-interacting case arises from the reduction of the Hilbert 
space of the system S in the limit $U\rightarrow\infty$; this influences the scattering phase shift of the 
tunneling electrons and quenches interference effects in this system due to a reduction of interfering tunneling pathways.

This change in the scattering phase shift has already been discussed by Wunsch et al. \cite{Wunsch2005} in terms of 
an interaction-induced renormalization of the (localized) orbitals (and was also found 
in more complex systems \cite{Darau2009,Donarini2010}). These results, however, have been derived assuming flat conduction bands, 
where additional renormalization effects due to the energy dependence of the conduction bands are not present. 
Thus, to provide deeper insight into the relevant physics, we compare results obtained from the HQME 
formalism (Sec.\ \ref{heomtheory}) to results obtained from the BM scheme 
(Sec.\ \ref{bornmarkovtheory}). The key difference between the two schemes is that the additional renormalization 
$\mathbf{\Delta}$ due to the coupling to the electrodes is included in the HQME method and discarded in the BM scheme. 
The BM approximation to the current-voltage characteristics and the real part of the coherence $\sigma_{10,01}$ are 
depicted by the dashed gray lines in Figs.\ \ref{model6current}a and \ref{model6currentX}, respectively.  As can be seen, 
the BM scheme yields overall larger current levels and gives, in particular, a much more rapid increase of the current 
level  at the onset of the resonant transport regime. The opposite holds true for the real part of the coherence $\sigma_{10,01}$. 
Considering the analysis at the end of Sec.\ \ref{desnonint} and that the level renormalizations $\Delta_{11}$ and $\Delta_{22}$  
are very similar due to the quasidegeneracy of the two levels,  we attribute the reduced current and decoherence levels 
obtained by the HQME scheme to the effect of the off-diagonal elements $\Delta_{12}$. Note that higher order effects have been ruled out 
by using lower truncation levels for the hierarchy (\ref{hierarcheom}).

The monotonic increase (decrease) of the current 
(coherence) level for bias voltages $\Phi>2\epsilon_{1/2}$ is also a consequence of the energy dependence of the level-width 
functions $\Gamma_{K,mn}$. They enter the Born-Markov master equations as $\Gamma_{K,mn}(\epsilon_{1/2})$. 
For bias voltages $e\Phi\gtrsim2\epsilon_{1/2}$, the value of $\Gamma_{\text{R/L},mn}(\epsilon_{1/2})$ decreases with an increasing 
bias voltage such that  destructive interference effects and the resulting current suppression become gradually less pronounced.

\begin{figure}
\begin{tabular}{l}
\hspace{-0.5cm}(a) \\
\resizebox{\newwidth}{\newheight}{
\includegraphics{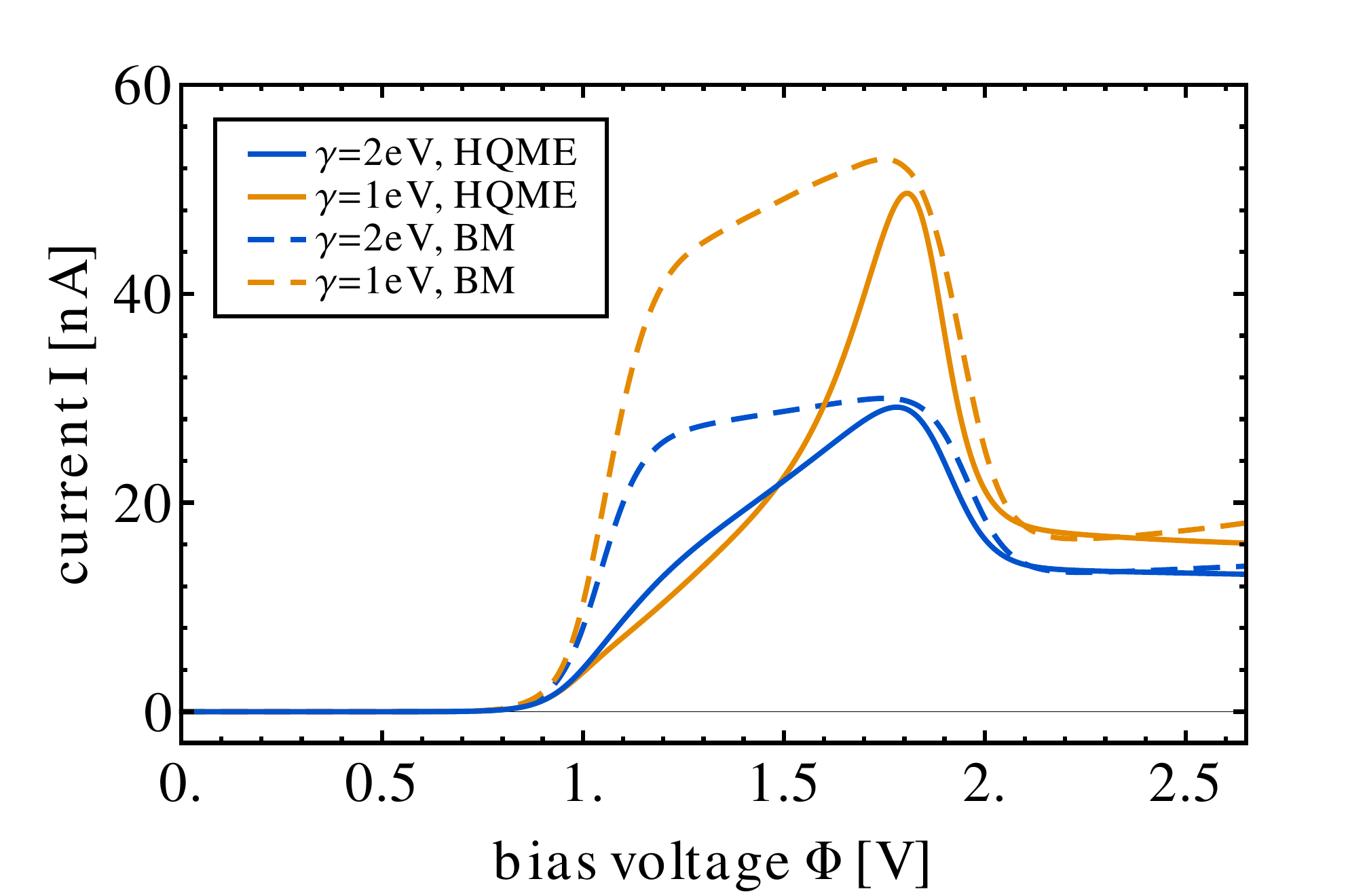}
}\\
\hspace{-0.5cm}(b) \\
\resizebox{\newwidth}{\newheight}{
\includegraphics{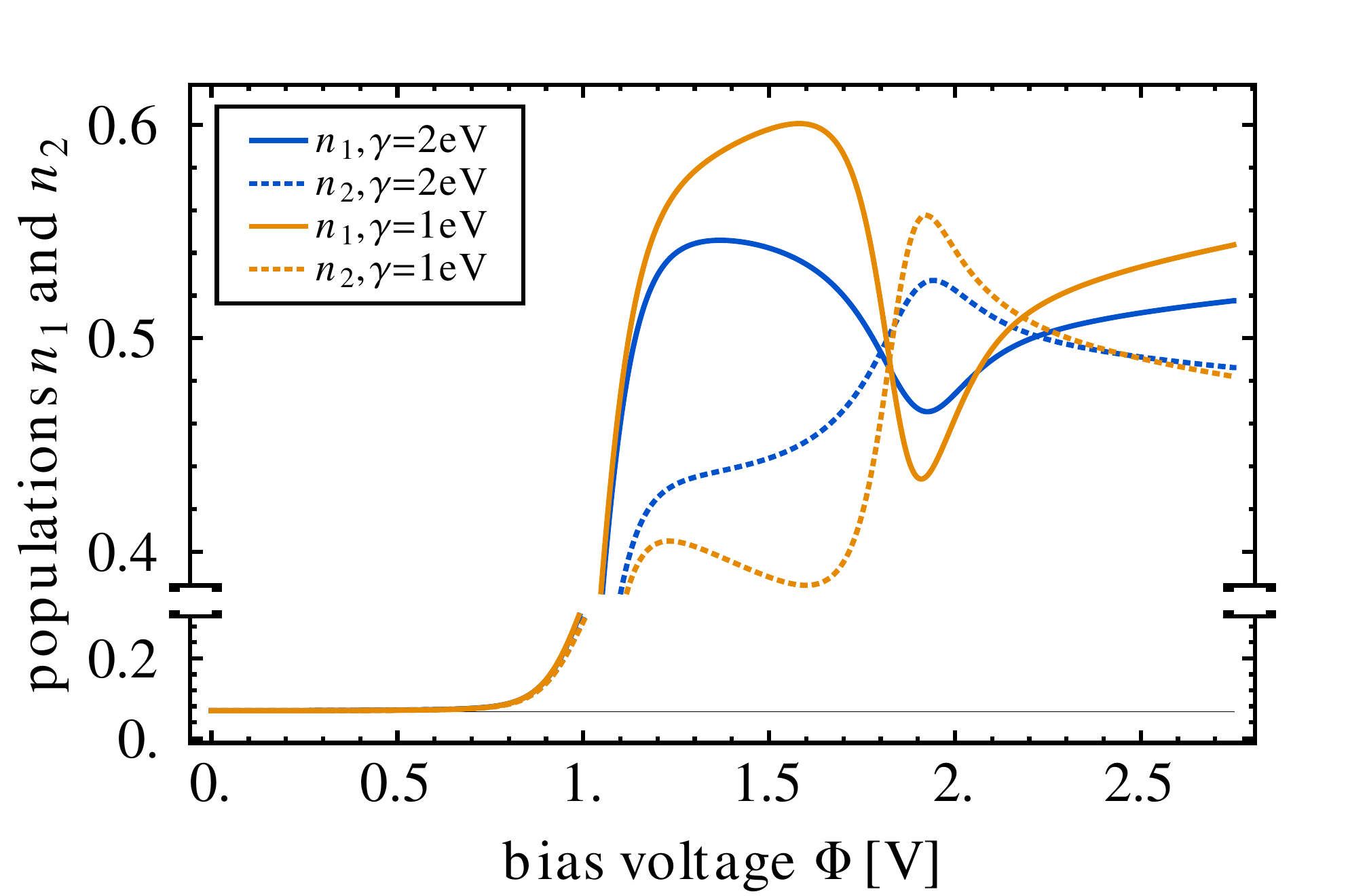}
}
\end{tabular}
\caption{(Color online) \label{model6band} Current-voltage and electronic population characteristics 
of junction DES for different band widths $\gamma$. The coupling parameter $\nu$ has been adjusted so that the maximal 
value of the level-width  functions $\Gamma_{K,mn}$ is the same in all cases. The solid and dotted lines have been 
obtained by the HQME method.  The dashed lines depict the corresponding current-voltage characteristics obtained  by the 
BM scheme.  Decoherence and lead induced inter-state coupling effects become stronger if the energy  dependence of the level-width functions 
$\Gamma_{K,mn}$ is more pronounced. 
}
\end{figure}

\subsubsection{Decoherence phenomena in the presence of finite electron-electron interaction strengths $U$}
\label{finUres}

In the  low and intermediate bias voltage regime $e\Phi<2\epsilon_{1/2}+U$ the finite $U$ case exhibits physics  similar 
to that of the infinite-$U$ case (see Fig.\ \ref{model6current}). One difference is that the system with finite electron-electron 
interactions has a  higher current while the difference in the electronic population is very similar. According to our analysis, 
the similarity in level occupation can be attributed to the off-diagonal elements $\Delta_{K,12}$ which have only a weak $U$ dependence. 
The associated current suppression, however, which arises from a modification of the effective level renormalization \cite{Wunsch2005} 
due to the effect of the $\Delta_{K,12}$, exhibits a non-negligible $U$-dependence. To corroborate this statement, the solid orange lines 
in Fig.\ \ref{model6band} show the transport characteristic of junction DES  calculated with a reduced band width $\gamma$, that 
is, effectively, with enhanced off-diagonal elements $\Delta_{K,12}(\epsilon)$. The current level of this junction is 
indeed reduced for bias voltages $e\Phi<2\epsilon_{1/2}+U$ while the difference in the electronic populations $n_{1}$ 
and $n_{2}$ is increased.

At higher bias voltages, $2\epsilon_{1/2}+U<e\Phi<2(\epsilon_{1/2}+U)$, the intermediate $U$ case exhibits a qualitatively 
different behavior from either the $U\rightarrow 0$ or $U\rightarrow \infty$ limits. The current of junction DES passes 
through a maximum value substantially larger than found in either of the limits  and then decreases, exhibiting a regime 
of negative differential resistance. The maximum value is seen to be close to the one found in the BM scheme 
(cf.\ Fig.\ \ref{model6current}a or Fig.\ \ref{model6band}a). Moreover, in the voltage regime of the current peak and negative 
differential resistance, an inversion of the electronic population occurs (see Fig.\ \ref{model6current}b). These effects 
are related to the energy dependence  of $\Gamma_{K,mn}(\epsilon)$  and arise when doubly occupied states enter the bias 
window.  As we have already seen in the limit $U\rightarrow\infty$, state $1$ is populated more strongly than state $2$ for $\Phi<2\epsilon_{1/2}+U$. 
Thus, at higher bias voltages, adding another electron into state $2$ is more favorable than into state $1$. 
Due to electron-electron interactions, these addition processes occur at higher energies $\epsilon_{1/2}+U$  and, therefore, 
also with a higher probability because of the energy dependence of the level-width functions:  
$\Gamma_{\text{L},mn}(\epsilon_{1/2}+U)>\Gamma_{\text{L},mn}(\epsilon_{1/2})$ for $e\Phi\approx2(\epsilon_{1/2}+U)$.  
When the population of the two levels is approximately the same, the effect of the off-diagonal elements $\Delta_{12}$ 
on the population dynamics is canceled. The associated current suppression also vanishes leaving the  decoherence due to electron-electron interactions 
as the dominant effect. The real part of the coherence $\sigma_{10,01}$ and the current level thus reach a local minimum and 
a maximum, respectively. Once the bias voltage exceeds $2(\epsilon_{1/2}+U)$, resonant transport through one of the levels 
occurs irrespective of the population of the other level.  Consequently, decoherence due to the reduction of the Hilbert 
space is no longer effective and the current level drops rapidly to the one of the non-interacting case.  Also, 
the population of the two levels start to follow again the same rules as in the non-interacting limit.

\begin{figure}
\resizebox{\newwidth}{\newheight}{
\includegraphics{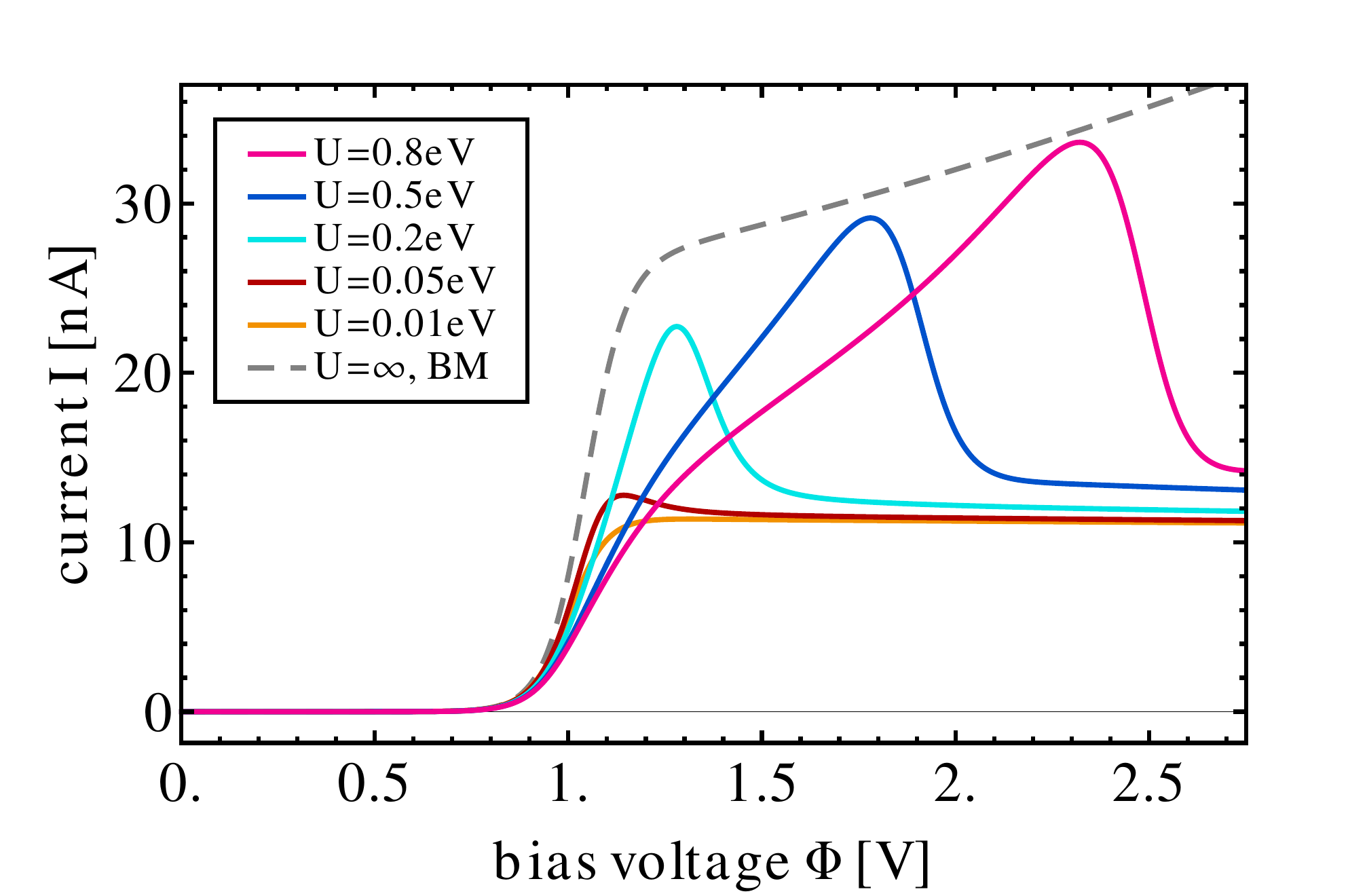}
}
\caption{(Color online)\label{model6current2} Current-voltage characteristics of junction DES 
for different electron-electron interaction strengths, ranging from $U=0.8$\,eV$\gg k_{\text{B}}T\approx25$\,meV 
to $U=0.01$\,eV$\ll k_{\text{B}}T$. Decoherence due to electron-electron interactions is independent 
of the interaction strength $U$, as long as it exceeds the broadening induced by the coupling to the electrodes and 
temperature. 
}
\end{figure}

The above analysis suggests that the maximal enhancement of the current due to decoherence by electron-electron interactions 
is not controlled by the value of the electron-electron interaction strength $U$. It rather enters via the asymmetry in the 
transfer rates $\Gamma_{\text{L},mn}(\epsilon_{1/2})$ and $\Gamma_{\text{R},mn}(\epsilon_{1/2})$ that increases with the 
applied bias voltage $\Phi$. This is confirmed by the current-voltage characteristics shown in Fig.\ \ref{model6current2} 
for different electron-electron interaction strengths $U$. 
As can be seen, the height of the current peaks for different values of $U$  follow the dashed gray line, which 
depicts the current-voltage characteristic  obtained from the BM scheme in the limit $U\rightarrow\infty$. Thereby, the similarity 
to the BM $U\rightarrow\infty$ case is only present if the electron-electron interaction strength of the system is 
significantly larger than the broadening of the steps in the corresponding current-voltage characteristic. 
If the broadening exceeds the electron-electron interaction strength $U$, which, in our case, is given by the thermal 
broadening $k_{\text{B}}T\approx25$\,meV, the decoherence effect becomes quenched. The same would be true if the broadening due to 
the coupling to the electrodes  is dominant or comparable to the thermal broadening, $\Gamma\gtrsim k_{\text{B}}T$,  as 
can be inferred from the results of Wunsch \emph{et al.}\ \cite{Wunsch2005}. This also means that our findings remain valid at 
higher temperatures, as long as $U\gg k_{\text{B}}T$.

\subsubsection{Asymmetric coupling to the electrodes}
\label{asymsec}

\begin{figure}
\begin{tabular}{l}
\hspace{-0.5cm}(a) \\
\resizebox{\newwidth}{\newheight}{
\includegraphics{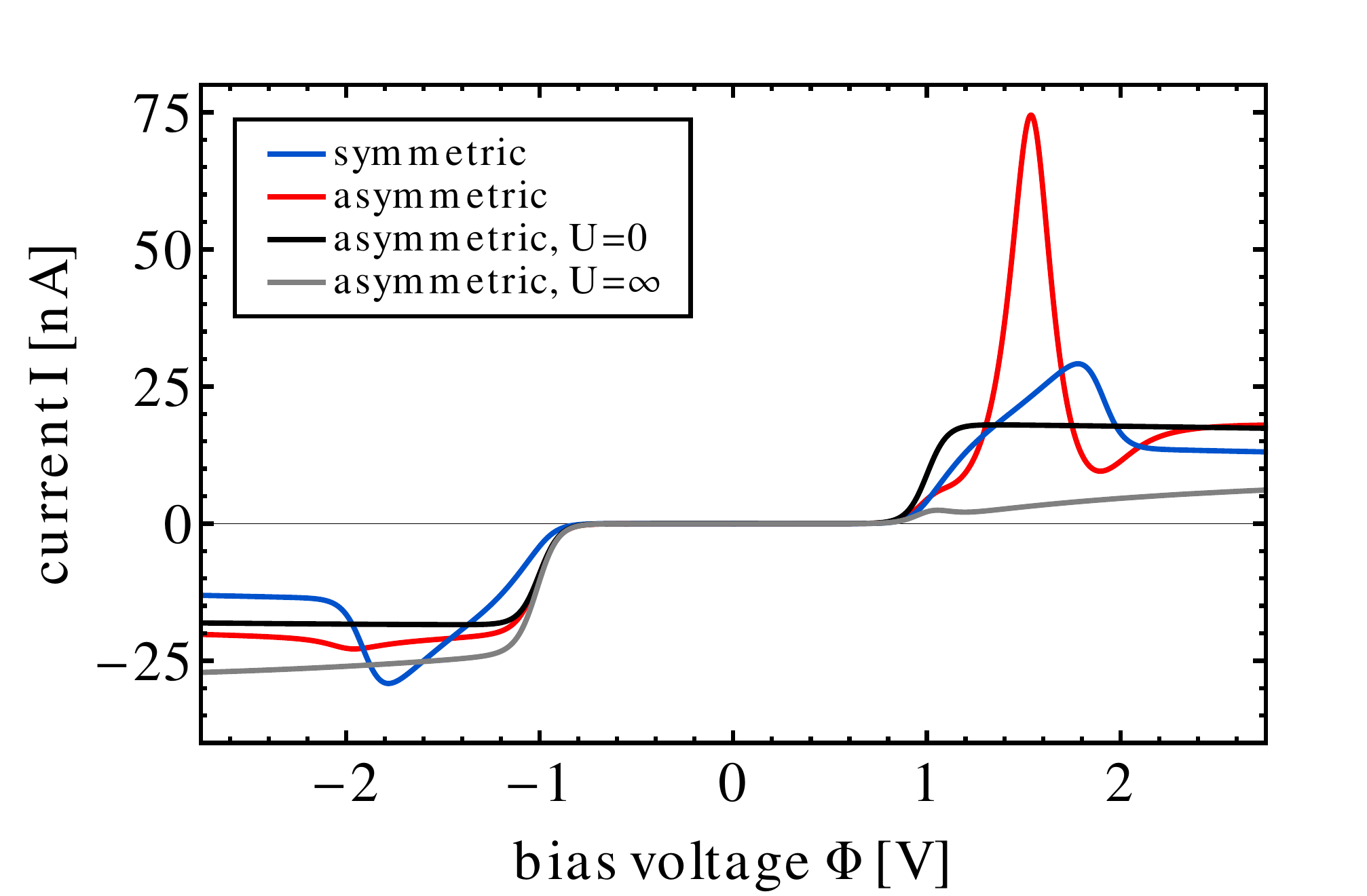}
}\\
\hspace{-0.5cm}(b) \\
\resizebox{\newwidth}{\newheight}{
\includegraphics{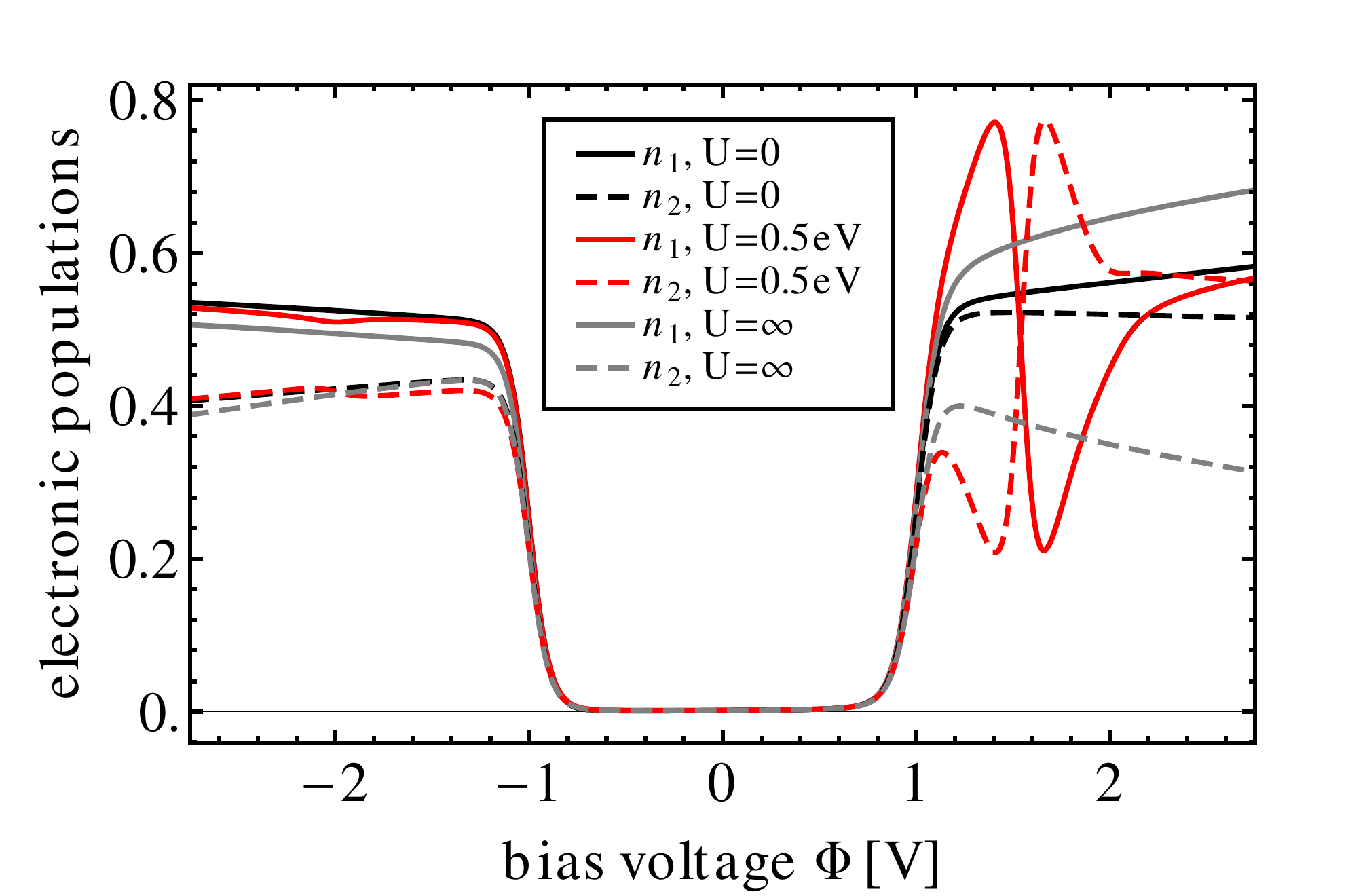}
}
\end{tabular}
\caption{(Color online) \label{model6asym} Panel (a): Current-voltage characteristics of the symmetrically coupled 
junction DES (solid blue line) and a very similar junction, where the coupling to the right lead is reduced by $1/3$ (red line). 
For comparison, the black and gray line show the current-voltage characteristics of the asymmetrically coupled system 
in the limit $U=0$ and $U=\infty$, respectively. 
Panel (b): Electronic population characteristics of the asymmetrically coupled system. 
Asymmetric coupling to the leads enhances the effect of population inversion 
and decoherence due to electron-electron interaction for one bias polarity, while for the other polarity 
both effects are attenuated. 
}
\end{figure}

So far, we considered only scenarios in which the absolute values of the coupling parameters $\vert\nu_{K,m}\vert$ 
are the same. In an experimental realization of junction DES the system S may be coupled more strongly to one of the 
electrodes than to the other. The current-voltage characteristic of such a junction is shown as the solid red line in 
Fig.\ \ref{model6asym}a. In comparison to junction DES (solid blue line), it is described by the same parameters 
except for a weaker coupling to the right electrode $\nu_{\text{R},1/2}=\nu/3$. 
The corresponding population of states 1 and 2 are depicted by the solid and dotted red lines in Fig.\ \ref{model6asym}b. 
As can be seen, an asymmetric coupling to the electrodes introduces a number of quantitative differences, including a 
particle-hole asymmetry and changes to the heights and widths of the current peaks, but the qualitative behavior is the 
same as in the symmetric case. It is interesting to note that similar current-voltage characteristics have been observed  
by Osorio \emph{et al.} \cite{Osorio2010} but only if the corresponding molecular junctions are in a low-current state.

\begin{figure}
\begin{tabular}{l}
\resizebox{\newwidth}{\newheight}{
\includegraphics{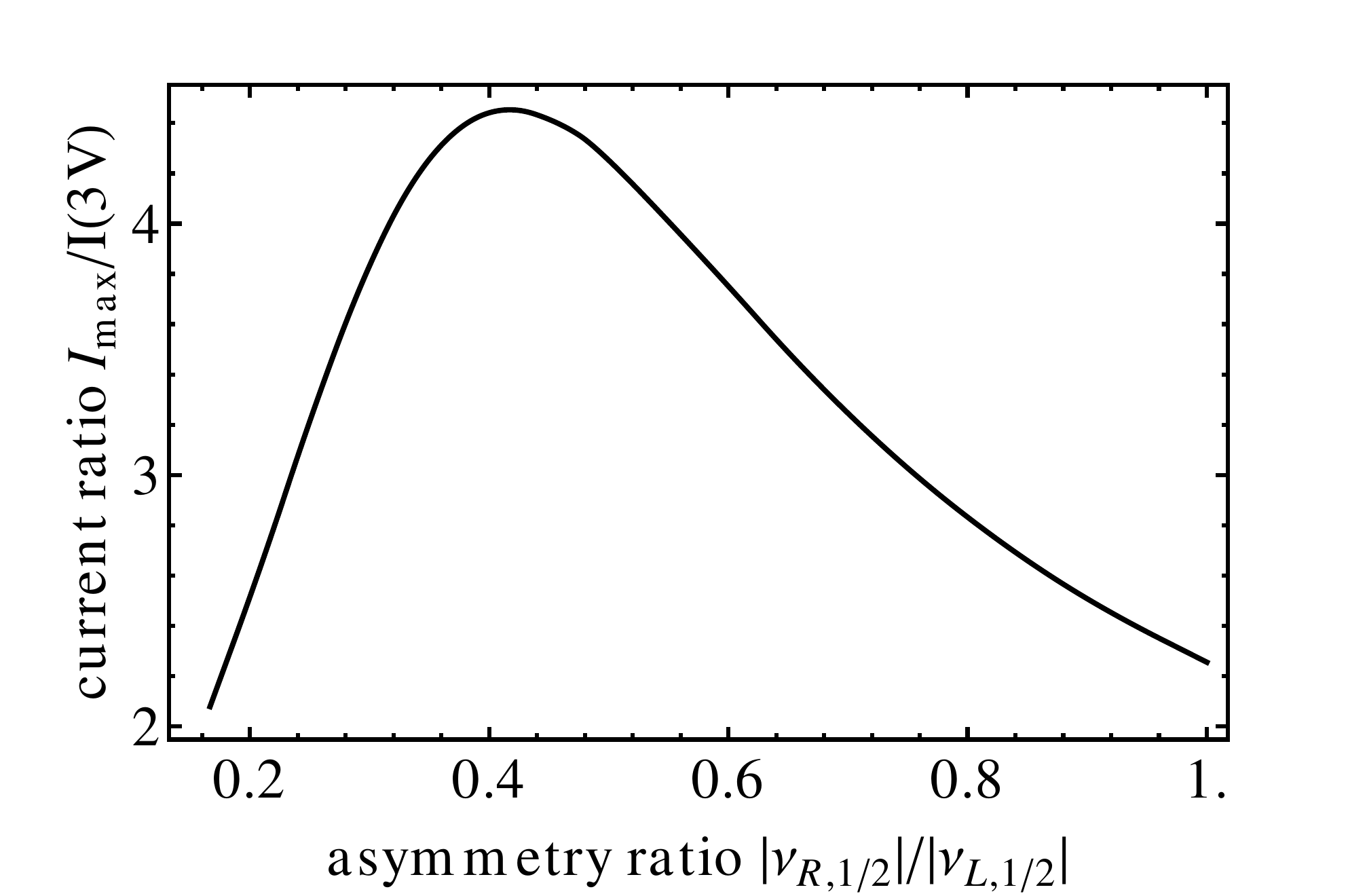} 
}
\end{tabular}
\caption{(Color online) \label{model6asym2} Ratio between the maximal current and the current level at $\Phi=3$\,V 
for junction DES as a function of the asymmetry ratio $\vert\nu_{\text{R},1/2}\vert/\vert\nu_{\text{L},1/2}\vert$. The peak in this curve 
reveals that 
decoherence due to electron-electron interactions is most effective for intermediate asymmetries in the system-electrode 
coupling
}
\end{figure}

The asymmetry with respect to bias voltage may be understood as follows. For positive bias voltages, where transport occurs 
from L$\rightarrow$S$\rightarrow$R, the two levels are populated on faster time scales than they become depopulated. 
Thus, the population of level $1$ is significantly larger at the onset of the resonant transport regime at $e\Phi=2\epsilon_{1/2}$. 
At this value of $\Phi$ the repulsive electron-electron interactions mean that the population of state $2$ is more strongly 
suppressed (cf.\ Fig.\ \ref{model6asym}b). 
The same arguments apply at higher bias voltages, $2\epsilon_{1/2}+U<e\Phi<2(\epsilon_{1/2}+U)$, where the population of the two 
levels become inverted once the doubly occupied states enter the bias window 
(cf.\ the discussion given in Sec.\ \ref{finUres}). Thus, the current suppression due to the blocking of transport channels  is 
more pronounced than for the symmetrically coupled junction such that, at $e\Phi\approx2\epsilon_{1/2}$ and $e\Phi\approx2(\epsilon_{1/2}+U)$,  
the current level drops to half of the level in the non-interacting case. 
This results also in a narrowing and a shift of the current peak to lower bias voltages. Compared to the symmetrically 
coupled scenario, the height of this peak is 
substantially increased, indicating that decoherence due to the reduced Hilbert space for the tunneling electrons is more pronounced. 
This is related to the reduced values of $\Gamma_{K,mn}$ in the same way as the increase of the current level in the limit 
$U\rightarrow\infty$ (cf.\ Sec.\ \ref{uinfsec}). Studying the ratio of the peak height versus the current level of the 
non-interacting case (which is equivalent to the current at $\Phi=3$\,V), one finds that this enhancement of the current peak 
is most pronounced for an asymmetry ratio 
$|\nu_{\text{R},1/2}|/|\nu_{\text{L},1/2}|\approx0.4$ (cf.\ Fig.\ \ref{model6asym2}).

At yet larger bias voltages, $e\Phi>2(\epsilon_{1/2}+U)$, destructive interference effects are fully developed and dominate the 
suppression of transport processes. The same holds true at negative bias polarities, 
where the time scales for populating and depopulating the two states are reversed such that electron-electron 
interactions and the corresponding decoherence mechanisms are, a priori, less pronounced. It is also interesting to note 
that, at these bias voltages, the current level of the asymmetrically coupled junction is slightly larger than for the symmetrically 
coupled one, despite the reduced coupling to the right lead. The physical origin of this behavior is that, again, destructive 
interference effects are less effective if the system is less strongly coupled to the electrodes \cite{Hartle2012}.

\subsubsection{Gate-voltage dependence} 
\label{gatesec}

Besides a source and a drain electrode, one often uses a third, so-called gate electrode to study the transport properties 
of a nanoelectronic device \cite{Tans97,Wegscheider2007,Ballmann2013}. Such an electrode is  capacitively coupled to the 
junction and acts to shift the level positions $\epsilon_{m}$ with respect to the Fermi levels of the leads. Dependence on 
gate voltage can help to reveal the nature of conduction processes.  Sequential tunneling  may be distinguished from 
higher-order processes such as cotunneling or spin-flip  (due to Kondo physics) \cite{Tews2004,Begemann2010} by its much stronger 
gate-voltage dependence. In the following, we study the gate voltage dependence of the transport properties of junction DES by 
assuming that the gate voltage shifts the level energies as $\epsilon_{m}\rightarrow\epsilon_{m}+e\Phi_{\text{gate}}$.

\begin{figure}
\begin{tabular}{ll}
(a)  & (b) \\
\resizebox{\newheight}{\newheight}{
\includegraphics{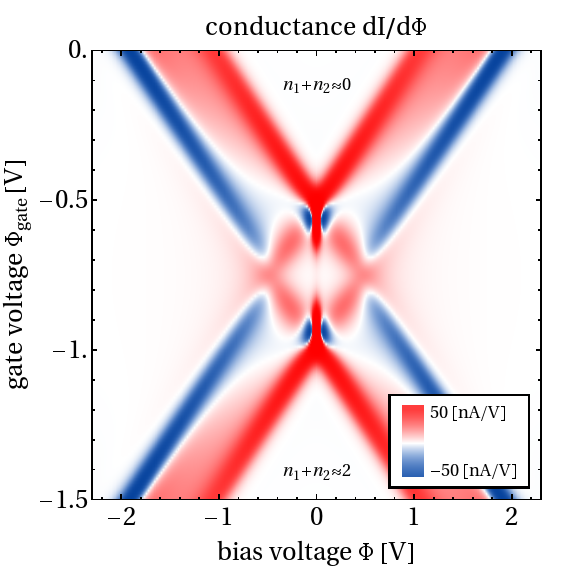}
}
&
\resizebox{\newheight}{\newheight}{
\includegraphics{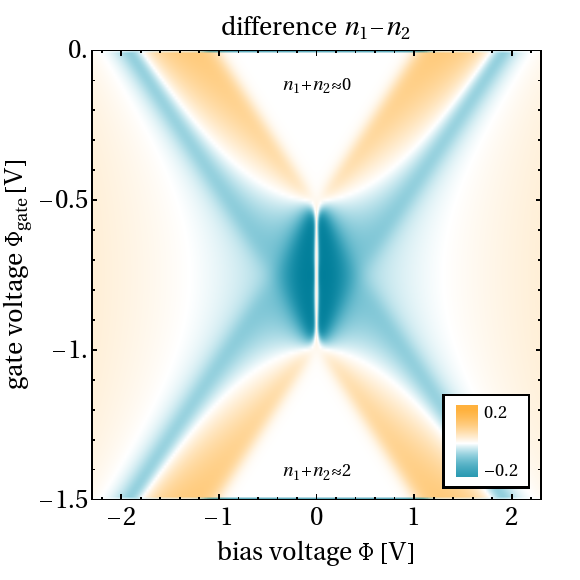}
}\\
\end{tabular}
\caption{(Color online) \label{maps} Panel (a) and (b) depict the differential conductance $\text{d}I/\text{d}\Phi$ 
and the difference in the electronic populations $n_{1}-n_{2}$ of junction DES as a function of both bias and gate voltage. 
Comparison shows that the occurrence of negative differential resistance (blue areas) is correlated with population inversion (turquoise areas) 
in the resonant transport regime. 
}
\end{figure}

Fig.\ \ref{maps}a presents a conductance map for junction DES in the plane of gate and source-drain voltage with negative 
conductance depicted in blue, zero in white and positive conductance in red.   The conductance map is symmetric with respect 
to both the bias voltage $\Phi$ and the gate-voltage $\Phi_{\text{gate}}$. The two symmetry axes cross at the charge-symmetric 
point $(e\Phi,e\Phi_{\text{gate}})=(0,-2\epsilon_{1/2}-U)\approx(0,-0.75$\,eV$)$. We therefore restrict the discussion to the 
upper right quarter of this map in the following. For gate-voltages $0>e\Phi_{\text{gate}}>-2\epsilon_{1/2}\approx-0.5$\,eV, 
the onset of the resonant transport regime at $e\Phi=2(\epsilon_{1/2}+\Phi_{\text{gate}})$ as well as the NDR feature 
appearing at  bias voltages slightly higher than the resonant tunneling onset decrease linearly  with the applied gate voltage. 
For yet smaller gate voltages, $-2\epsilon_{1/2}>e\Phi_{\text{gate}}>-2\epsilon_{1/2}-U$ where $\epsilon_{1/2}+e\Phi_{\text{gate}}<0$, 
this trend is reversed. In addition, a more complex structure emerges, as the current drops in two steps at 
$e\Phi\gtrsim-2(\epsilon_{1/2}+e\Phi_{\text{gate}})$ and $e\Phi\lesssim2(\epsilon_{1/2}+U+\Phi_{\text{gate}})$. 
The current maximum and the corresponding NDR vanish as the system is driven to the charge-symmetric point 
(at $e\Phi_{\text{gate}}=-2\epsilon_{1/2}-U$).

\begin{figure}
\resizebox{\newheight}{\newheight}{
\includegraphics{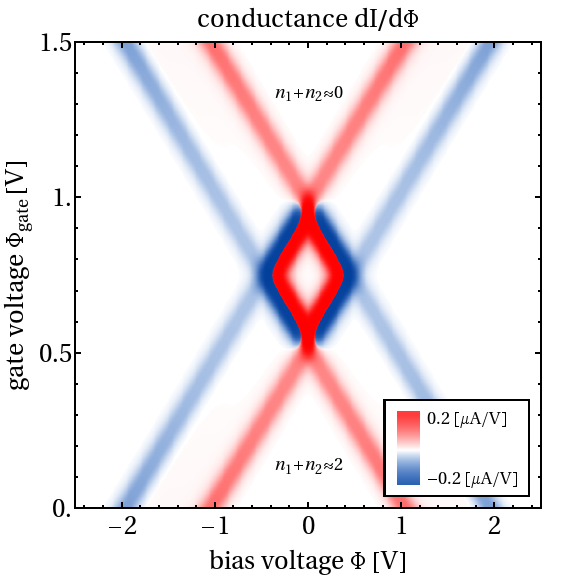}
}
\caption{(Color online) \label{maps2} Conductance map similar to the one shown in Fig.\ \ref{maps}a  
but obtained using the BM master equation scheme (cf.\ Sec.\ \ref{bornmarkovtheory}). Comparison of Figs.\ \ref{maps}a 
and \ref{maps2} reveals the strong renormalization effects that are induced by lead induced inter-state coupling encoded in 
the off-diagonal elements $\Delta_{12}$. 
}
\end{figure}

The nature of the additional drop in the current can be revealed by inspection of the conductance map obtained from the 
BM master equation scheme shown in Fig.\ \ref{maps2}. In the range of gate voltages from $-2\epsilon_{1/2}$ to $-2\epsilon_{1/2}-U$, 
the onset of the resonant transport regime at $e\Phi=-2(\epsilon_{1/2}+e\Phi_{\text{gate}})$ does not appear as a step but 
as a peak in the respective current-voltage characteristic. This is reflected by high positive conductance values for 
$e\Phi\lesssim-2(\epsilon_{1/2}+e\Phi_{\text{gate}})$ followed by negative values 
for $e\Phi\gtrsim-2(\epsilon_{1/2}+e\Phi_{\text{gate}})$. The consecutive drop of the 
current at $e\Phi\sim2(\epsilon_{1/2}+U+e\Phi_{\text{gate}})$ is almost the same as for higher gate voltages, 
$0>e\Phi_{\text{gate}}>-2\epsilon_{1/2}$.  
The appearance of the current peaks is related to the fact that, for $e\Phi\lesssim-2(\epsilon_{1/2}+e\Phi_{\text{gate}})$, 
transport processes occur via a thermally assisted 
tunneling process from the junction into the right lead, followed by another tunneling process from the left lead onto the junction. 
While the former exhibits partial Pauli-blocking, the latter does not and, therefore, involves electrons with a wider range of energies. 
As a result, destructive interference effects cannot fully develop. This results in a more pronounced increase of the current level than for 
thermally assisted transport through the unoccupied (or doubly occupied) system. For larger bias voltages, 
$e\Phi\gtrsim-2(\epsilon_{1/2}+e\Phi_{\text{gate}})$, 
the Pauli-blocking of the tunneling process into the right electrode is no longer active and destructive interference effects start 
to suppress the current flow through the junction, reducing the current level again.

Considering again Fig.\ \ref{maps}a, where, in contrast to Fig.\ \ref{maps2}, the real part of the 
self-energy matrix $\mathbf{\Sigma}$ is taken into account, one observes a strong renormalization of the 
above described temperature-induced decoherence effects, in particular of the current peaks at 
$e\Phi\gtrsim-2(\epsilon_{1/2}+e\Phi_{\text{gate}})$. On one hand, they become quenched the closer the system is driven to the 
charge-symmetric point (\emph{i.e.} for gate voltages $\Phi_{\text{gate}}\approx-(\epsilon_{1/2}+U/2)=-0.75$\,V). 
Thereby, it should be noted that the renormalization of these peaks leads to signatures around zero bias that appear 
to be similar (but must not be mistaken) as signatures due to pseudo-Kondo \cite{Hofstatter2001,Lee2007} or Kondo-like 
correlations \cite{Cronenwett1998,Goldhaber1998,Liang02}. 
On the other hand, these peaks appear much broader. While this broadening exceeds the thermal broadening ($\approx25$\,meV) or 
the broadening due to the coupling to the electrodes ($\approx2$\,meV) by more than an order of magnitude, it is most pronounced 
around the charge-symmetric line, \emph{i.e.} for $\Phi_{\text{gate}}\approx-0.75$\,V, where it amounts to $\sim0.5$\,eV. 
The other (electronic) signatures of the conductance map exhibit similar broadening. 
Note that such a pronounced broadening of electronic signatures has recently been experimentally observed in the transport 
characteristics of a number of single-molecule junctions \cite{Secker2010}, where, however, it has been attributed to 
electronic-vibrational coupling \cite{Ness05,Secker2010}.

Fig.\ \ref{maps}b shows the difference in the electronic populations $n_{1}$ and $n_{2}$.  Comparison to Fig.\ \ref{maps}a  
reveals that the appearance of negative differential resistance is closely linked to population inversion (turquoise areas) 
in the resonant transport regime. In the non-resonant transport regime and gate voltages 
$-2\epsilon_{1/2}>e\Phi_{\text{gate}}>-2\epsilon_{1/2}-U$, the population of the higher lying level is also more pronounced, 
although the current is monotonously increasing. As this behavior is not observed using the Born-Markov scheme, we also 
attribute it to the inter-state coupling effects induced by the off-diagonal elements $\Delta_{12}$. Note that the corresponding 
relaxation time scale is three orders of magnitude larger in this regime (ns instead of ps). We therefore computed the 
detailed data shown in Figs.\ \ref{maps}a and \ref{maps}b by truncating the hierarchy of equations of motion (\ref{hierarcheom}) 
at the first tier. We checked that the results that have been discussed in this section are not affected by this choice of 
the truncation scheme.

\subsubsection{Comparison to blocking state scenarios}
\label{compblock}

The mechanism for negative differential resistance that we investigate in this article is based on decoherence 
due to electron-electron interactions. Another well-known mechanism for NDR due to electron-electron interactions involves 
electronic states that are weakly coupled to the electrodes. This includes blocking states 
\cite{Hettler2002,Hettler2003,Datta2007,Hartle2010b,Leijnse2011} which are weakly coupled to only one of the electrodes as 
well as centrally localized states \cite{Hartle2010b} which are weakly coupled to both electrodes (or, similarly, spin-blockade 
in systems that are coupled to ferromagnetic leads \cite{Braun2004}). In these systems, 
NDR occurs when the weakly coupled state enters the bias window, leading to similar features in the current-voltage 
characteristic than the decoherence mechanism outlined before. Inspection of the respective 
conductance maps, however, reveals important qualitative differences.

\begin{figure}
\begin{tabular}{ll}
a)  & b) \\
\resizebox{\newheight}{\newheight}{
\includegraphics{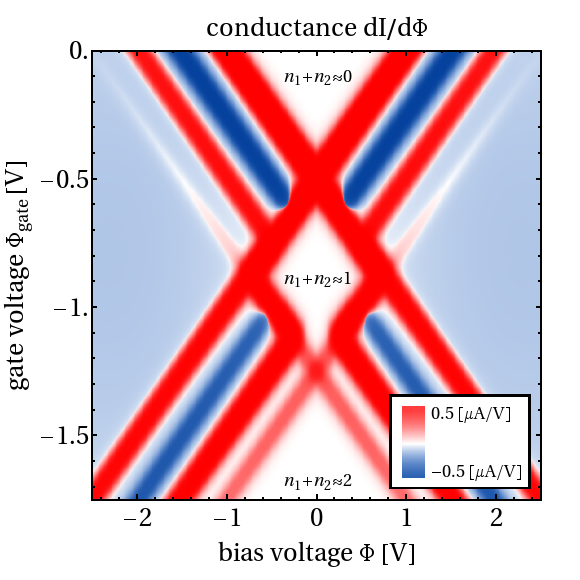}
}
&
\resizebox{\newheight}{\newheight}{
\includegraphics{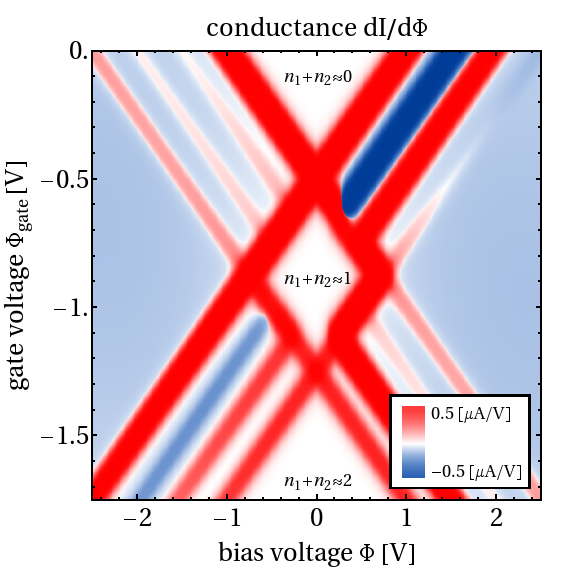}
}\\
\end{tabular}
\caption{(Color online) \label{maps3} Conductance maps of junctions with a centrally localized state (Panel (a)) 
and a blocking state (Panel (b)). 
}
\end{figure}

Fig.\ \ref{maps3}a and \ref{maps3}b show conductance maps of junctions with a centrally localized (model CENTRAL) 
and a blocking state, respectively (model BLOCK). A detailed list of the corresponding model parameters is found in 
Tab.\ \ref{parameters}.  The most important difference between NDR due to decoherence and NDR due to weakly coupled states 
is the symmetry with respect to the applied gate voltage. While decoherence leads to symmetric NDR features, weakly coupled 
states result in NDR that is non-symmetric with respect to the gate voltage $\Phi_{\text{gate}}$. Moreover, areas where NDR 
occurs terminate in different ways at the central Coulomb diamond. While these areas almost touch each other when NDR is 
induced by decoherence, they terminate at the sides of the diamond when NDR is the result of weakly coupled states. Note 
that an asymmetric coupling to one of the electrodes, for example a weaker coupling to the right electrode, 
attenuates the NDR features only in the upper left and the lower right corner (similar as in Fig.\ \ref{maps3}b), irrespective 
of the underlying mechanism (data not shown).

\subsubsection{Influence of the type of conduction bands: Comparison to NCA}
\label{ncaresults}

\begin{figure}
\resizebox{\newwidth}{\newheight}{
\includegraphics{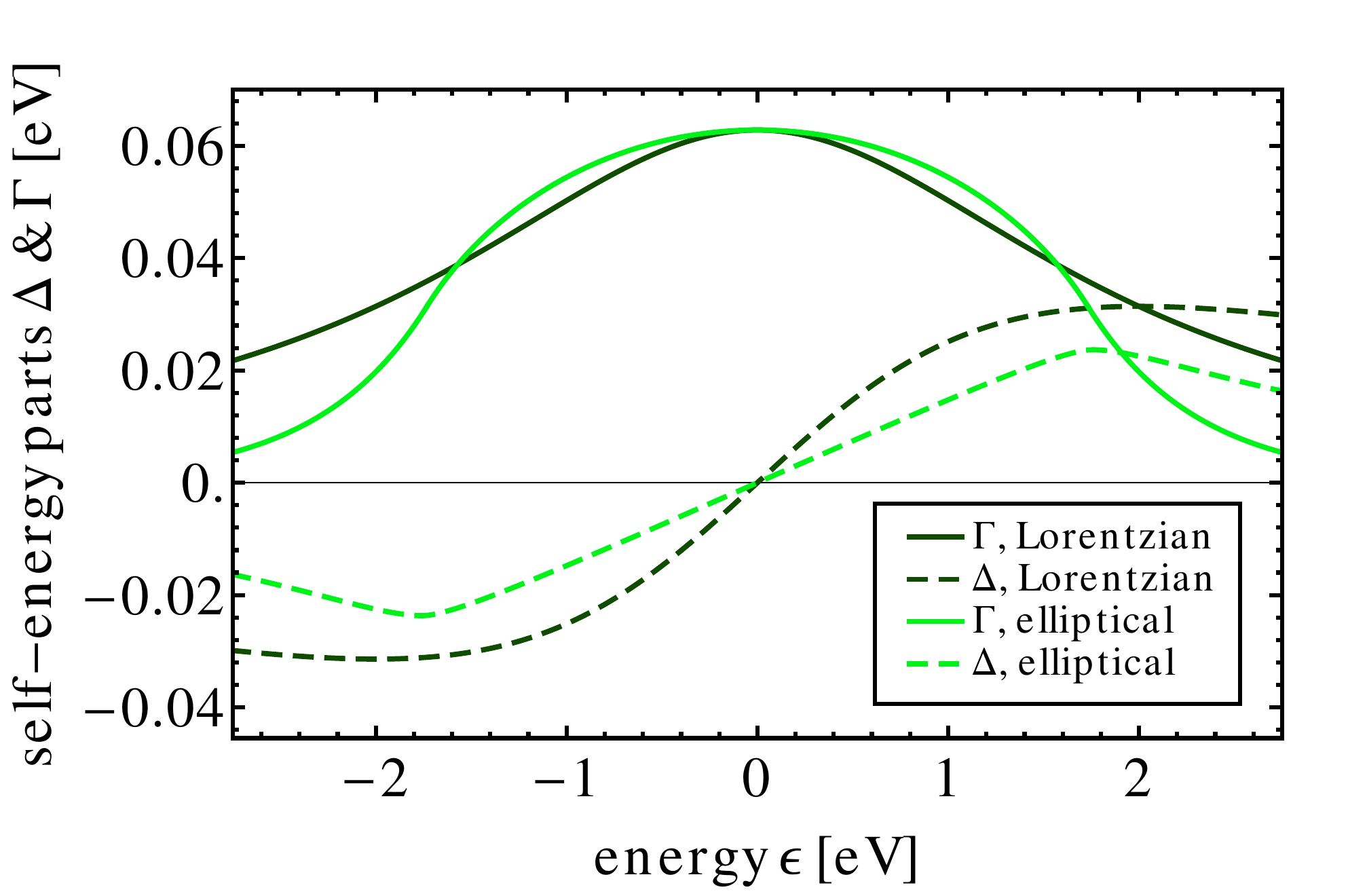}
} 
\caption{(Color online)\label{semiband} Level-width functions and the corresponding real parts of the self 
energy as a function of energy $\epsilon$. The dark green lines refer to Lorentzian conduction bands (cf.\ Eq.\ \ref{levelwidths}), 
while the light green lines are associated with the (modified) semi-elliptical bands (cf.\ Eq.\ (\ref{semielleq})). 
}
\end{figure}

In this section we use the NCA (cf.\ Sec.\ \ref{ncatheory}) to investigate the interplay 
between decoherence and correlation in junction DES. 
The NCA is an infinite order resummation of a selected subset of diagrams in a 
hybridization expansion. As will be seen, it reproduces some of the qualitative behavior that is found by the numerically exact 
HQME method, although it also misses other aspects and is quantitatively inaccurate. An advantage, however, is that it facilitates the study of 
non-Lorentzian density of states. As an example, we present here results obtained using  semi-elliptical conduction bands   
\begin{equation}
\label{semielleq}
\Gamma_{K,mn}\left(\epsilon\right)=2\pi\frac{\nu_{K,m}\nu_{K,n}}{\gamma}\sqrt{1-\frac{\epsilon^{2}}{\gamma^{2}}}, 
\end{equation}
For numerical convenience we slightly modified the functional dependence at the band edges, replacing the square root 
singularity by an exponential decay, so $\Gamma_{K,mn}(\epsilon)=Ae^{-\alpha\left|\epsilon-\mu_{K}\right|}$ when 
$\vert\epsilon-\mu_{K}\vert>\epsilon_{c}$,   where $\Gamma_{K,mn}(\epsilon_{c})=0.5 \Gamma_{K,mn}(\mu_{K})$ and the 
parameters $A$ and $\alpha$ are chosen such that $\Gamma_{K,mn}$ and its first derivative are continuous. The resulting 
level-width function(s) and the respective real parts of the self energy are depicted in Fig.\ \ref{semiband}.

\begin{figure}
\begin{tabular}{l}
\hspace{-0.5cm}(a) \\
\resizebox{\newwidth}{\newheight}{
\includegraphics{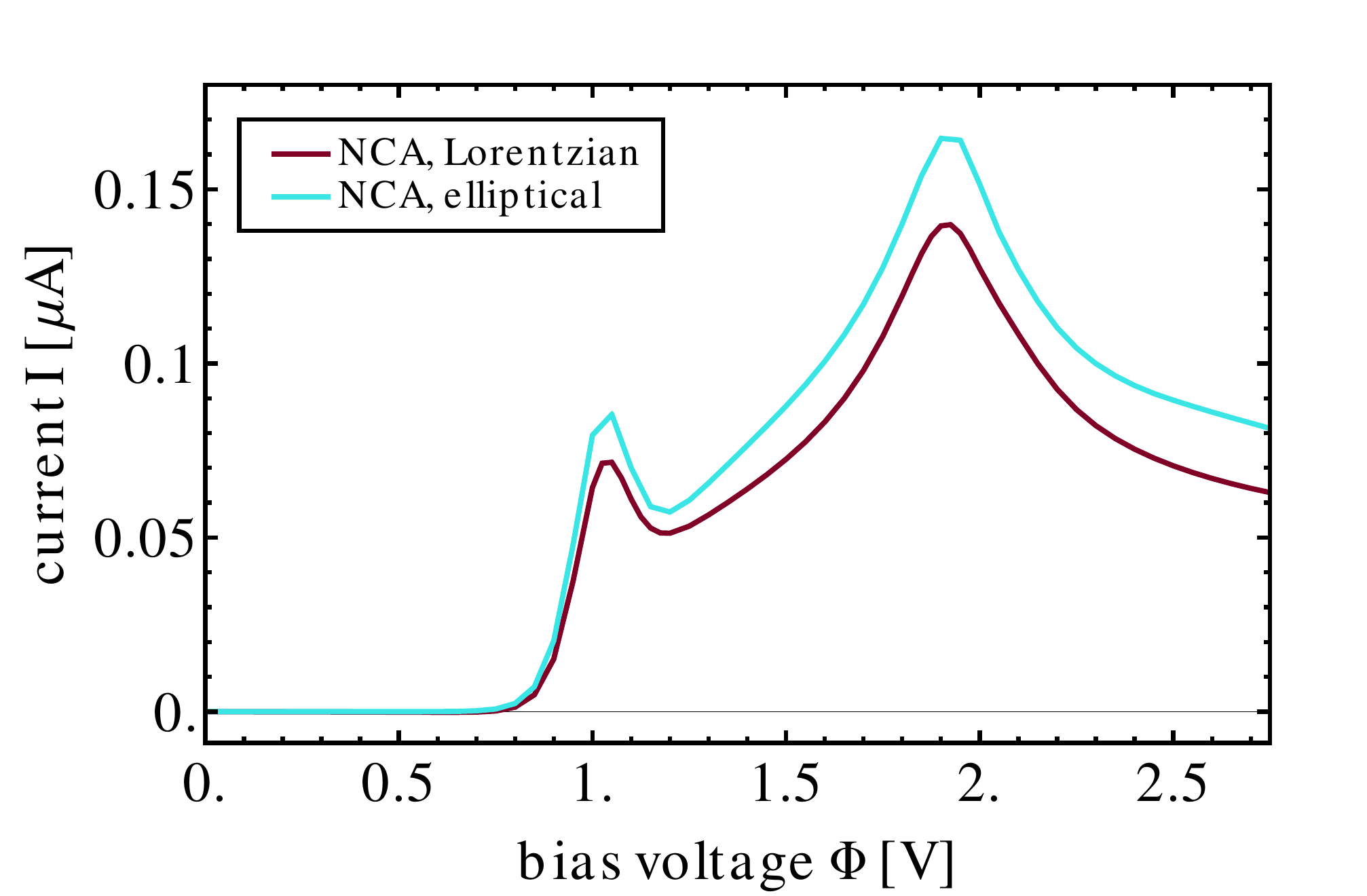}
} \\
\hspace{-0.5cm}(b) \\
\resizebox{\newwidth}{\newheight}{
\includegraphics{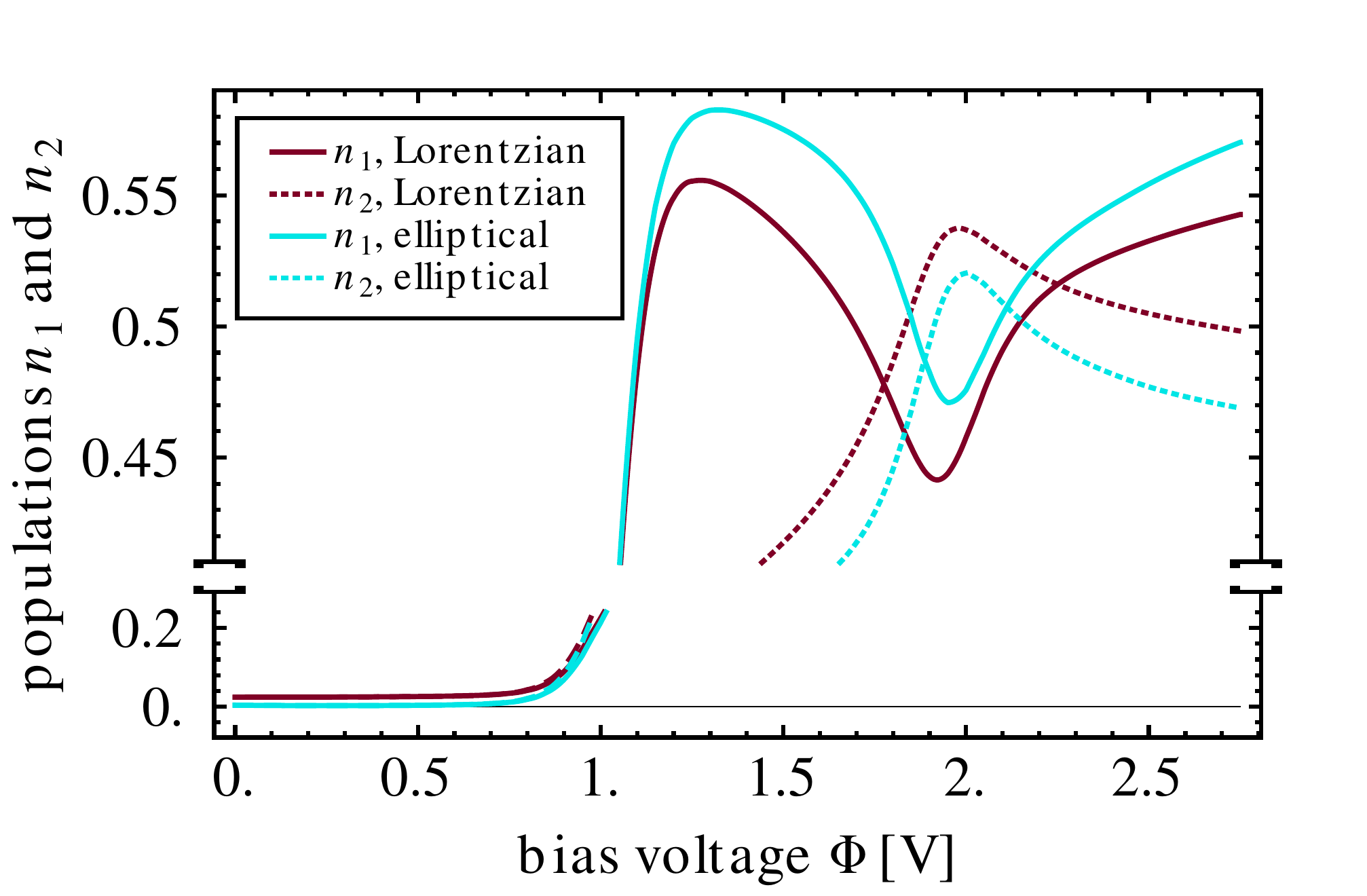}
} \\
\end{tabular}
\caption{(Color online)\label{model6nca} Current-voltage and electronic population characteristics of junction DES. 
The turquoise and red curves depict results that are obtained with the NCA scheme 
for elliptical and Lorentzian conduction bands, respectively. 
}
\end{figure}

The corresponding current-voltage and electronic population characteristics are represented by the turquoise lines in 
Fig.\ \ref{model6nca}. For comparison, the dark red lines show results that have been computed using Lorentzian conduction 
bands (cf.\ Eq.\ (\ref{levelwidths})). The basic decoherence phenomena found for the Lorentzian density of 
states also occur for elliptical conduction bands, in particular the inter-state coupling effects encoded in the off-diagonal 
elements $\Delta_{12}$, that is an enhanced broadening of electronic signatures and population inversion. 
While the NCA captures these effects qualitatively, it is also evident that it fails on both a quantitative 
and a qualitative level. According to our previous analysis, for example, the differences in the electronic population 
should be less pronounced because the real parts of the corresponding self-energy matrix are smaller (cf.\ Fig.\ \ref{semiband}). 
This picture can be confirmed very easily in the non-interacting case, using, for example, the Green's function analysis 
outlined in Sec.\ \ref{desnonint}. However, it is reproduced by the NCA results only in parts. In particular, the difference of the 
electronic populations at the onset of the resonant transport regime $e\Phi\approx2\epsilon_{1/2}$ and for higher bias 
voltages $e\Phi>2(\epsilon_{1/2}+U)$ is not smaller but larger in the elliptical case. 
This already indicates that the partial resummation of the NCA diagrams is not sufficient to describe this transport problem.

\begin{figure}
\resizebox{\newwidth}{\newheight}{
\includegraphics{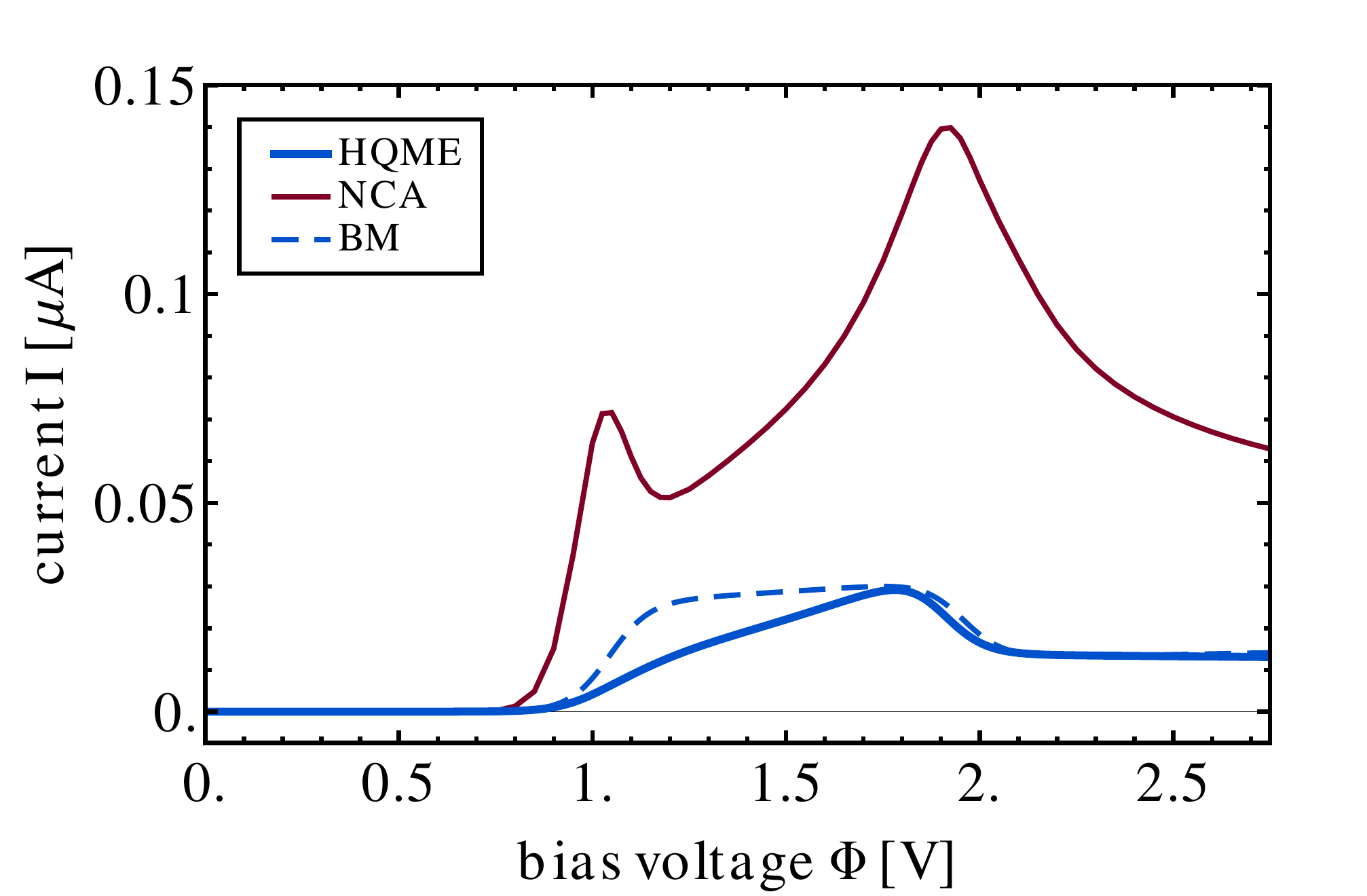}
} 
\caption{(Color online)\label{model6ncacompmethod} Current-voltage characteristics of junction DES obtained by three different methods: 
the HQME method (blue line), the NCA (dark red line) and the Born-Markov scheme (dashed blue line). 
}
\end{figure}

A direct comparison of the NCA results with the ones obtained by the BM and the HQME approach reveals further inaccuracies. 
To this end, we show the current-voltage characteristics of junction DES in Fig.\ \ref{model6ncacompmethod} that have been 
obtained by the three methods using Lorentzian conduction bands. 
Again, while the NCA captures some aspects of the problem, like the decoherence-induced negative differential resistance, 
it also shows qualitative differences. This includes an additional current peak at $e\Phi\approx2\epsilon_{1/2}$ but, 
more importantly, also much higher current levels. Thus, NCA underestimates the effect of destructive interference effects 
in this system. Future work may show if these deficiencies can be overcome using the one-crossing (or higher) approximation  
in order to obtain qualitatively or even quantitatively correct results \cite{Grewe2008}. Note that 
another nonequilibrium example where NCA fails in the presence of non-degenerate levels is given in Ref.\ \cite{Cohen2013}.

\subsubsection{Higher order effects} 
\label{numerisec}

\begin{figure}
\begin{tabular}{l}
\resizebox{\newwidth}{\newheight}{
\includegraphics{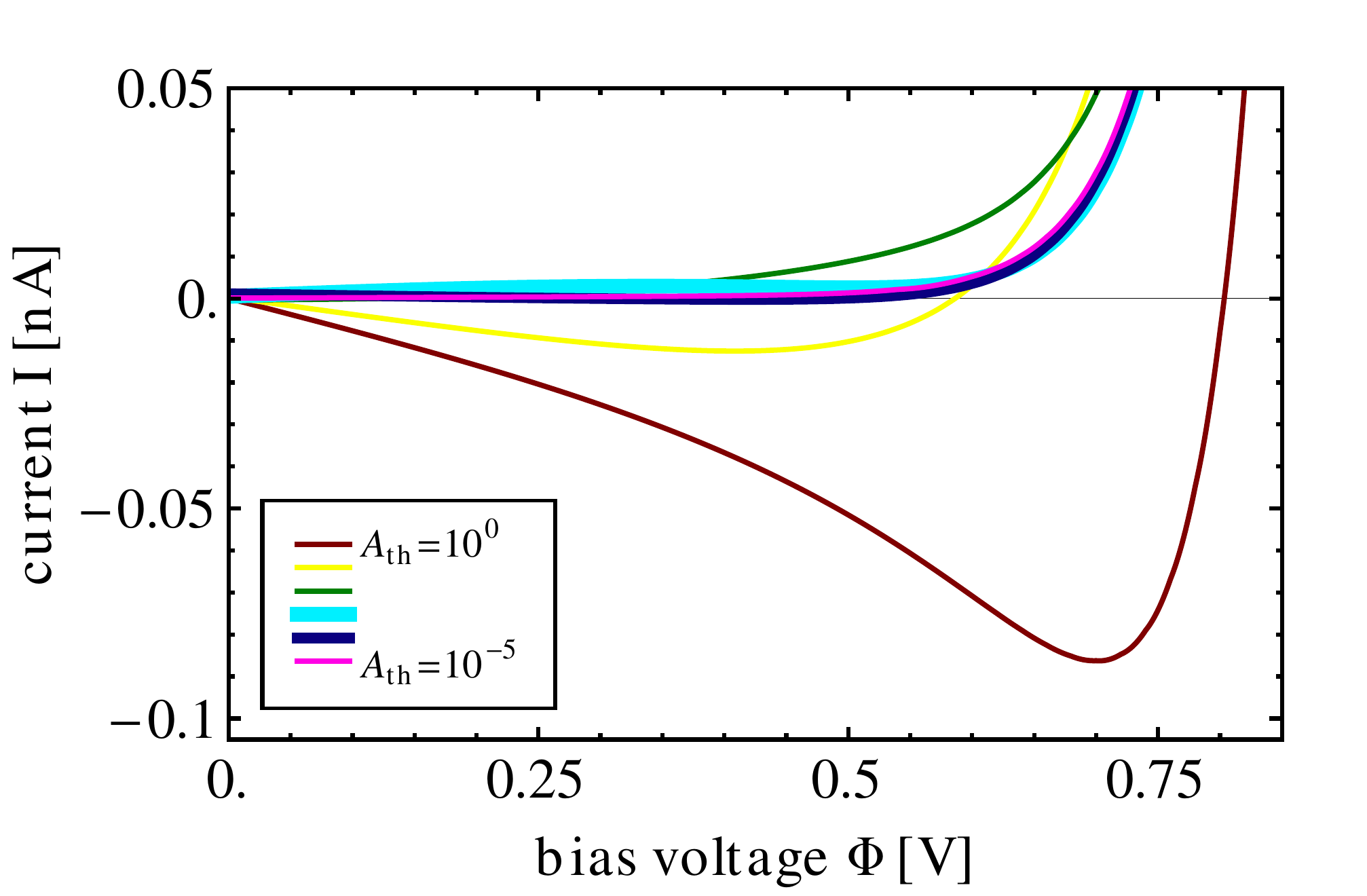}
}
\end{tabular}
\caption{(Color online) \label{model6num} Current-voltage characteristics of junction DES for different depths 
of the hierarchy of auxiliary operators (controlled by the threshold value $A_{\text{th}}$). At the lowest level, 
only resonant processes are included, while, upon increasing the depth of the hierarchy, higher order processes enter 
successively. Unphysical results are obtained in the non-resonant transport regime, \emph{i.e.}\ for 
$e\Phi<2\epsilon_{1/2}\approx1$\,V, 
if the depth of the hierarchy is too low. 
}
\end{figure}

At this point, some remarks about higher order effects and numerics are in order. Fig.\ \ref{model6num} presents  
current-voltage characteristics for junction DES, employing different truncation levels of the hierarchy (\ref{hierarcheom}). 
Thereby, we focus on the non-resonant transport regime, $e\Phi<2\epsilon_{1/2}$, where resonant processes are energetically not accessible 
or Pauli-blocked and, consequently, higher order effects are dominant. This requires a theoretical description beyond 
the leading order in $\Gamma$ \cite{,Hettler2007,Leijnse09,Esposito2010}. In the HQME methodolody, the depth of such an expansion can be controlled by the 
threshold value $A_{\text{th}}$, where lower values correspond to a higher quality of the result. The details of 
the truncation scheme are outlined in App.\ \ref{appB}. Tab.\ \ref{tabnum} summarizes the number of 
auxiliary operators $\sigma^{(\alpha)}_{j_{1}..j_{\alpha}}$ that are included in these calculations at every tier. 
The number of auxiliary operators in the $\alpha$th tier can be estimated as $\gamma^{\alpha-1}/((k_{\text{B}}T)^{\alpha-1}A_{\text{th}}\alpha!)$ 
(cf.\ App.\ \ref{appB}). In the resonant transport regime, we observe only marginal variations in the transport characteristics of junction DES 
as the truncation level is changed, because the first tier, including the effect of resonant processes, is always fully accounted for 
by our scheme.

\begin{table}
\begin{center}
\begin{tabular}{|c|*{8}{ccc|}}
\hline 
\backslashbox{~~$\alpha$}{$A_{\text{th}}$~} && $10^{0}$&&&$10^{-1}$&&& $10^{-2}$&&&$10^{-3}$&&&$10^{-4}$&&&
$10^{-5}$&\\ \hline 
 1 && 202  &&& 202  &&& 202  &&& 202  &&& 202  &&& 202  &\\
 2 && 0  &&& 12  &&& 97  &&& 423  &&& 1530  &&& 3486  &\\
 3 && 0  &&& 0    &&& 0   &&& 17  &&& 311  &&& 1965 &\\
 4 && 0  &&& 0    &&& 0   &&& 0  &&& 0  &&& 7 &\\
\hline 
\end{tabular}
\end{center}
\caption{\label{tabnum} Number of auxiliary operators in each tier $\alpha$ 
for different threshold values $A_{\text{th}}$. Thereby, the first hundred Matsubara frequencies $\omega^{s}_{K,1..100}$ 
are included in order to obtain converged results. 
}
\end{table}

\begin{table}
\begin{center}
\begin{tabular}{|c|*{8}{ccc|}}
\hline 
\backslashbox{~~$\alpha$}{$A_{\text{th}}$~} && $10^{0}$&&&$10^{-1}$&&& $10^{-2}$&&&$10^{-3}$&&&$10^{-4}$&&&
$10^{-5}$&\\ \hline 
 1 && 62  &&& 62  &&& 62  &&& 62  &&& 62  &&& 62  &\\
 2 && 0  &&& 10  &&& 85  &&& 392  &&& 1381  &&& 2896  &\\
 3 && 0  &&& 0    &&& 0   &&& 12  &&& 235  &&& 1442 &\\
 4 && 0  &&& 0    &&& 0   &&& 0  &&& 0  &&& 3 &\\
\hline 
\end{tabular}
\end{center}
\caption{\label{tabnumpade} Number of auxiliary operators in each tier $\alpha$ for different threshold values $A_{\text{th}}$ 
if, instead of the Matsubara decomposition scheme, the Pade approximation \cite{Ozaki2007,Hu2010,Hu2011} is employed. 
Thereby, we included the first thirty poles of this decomposition scheme in order to obtain converged results. 
}
\end{table}

The results are converged to a satisfactory level if $A_{\text{th}}\lesssim10^{-5}$. 
They are the same as if all auxiliary operators of the zeroth, 
first and second tier are included while all other auxiliary operators are discarded (data not shown). 
Unphysical negative currents are obtained for higher threshold values $A_{\text{th}}\gtrsim10^{-2}$, which corresponds 
to the same level of accuracy as the BM master equation scheme including, however, the real part of the self-energy due 
to the coupling to the electrodes. 
This shows both the importance of second order effects, which are only well represented 
for $A_{\text{th}}\lesssim10^{-4}$, and the stability of our results with respect to higher order effects. Decoherence due 
to electron-electron interactions is less pronounced in the non-resonant transport regime, because the population 
of the two levels is negligible, $n_{1/2}\approx0$.

For comparison we also show the number of auxiliary operators in Tab.\ \ref{tabnumpade} that would have been required 
using the alternative Pade approximation scheme \cite{Ozaki2007,Hu2010,Hu2011}. The results are almost identical to the 
ones shown in Fig.\ \ref{model6num} (data not shown). Although the Pade approximation is, in general, more efficient, the increase  is 
limited. In the present case, the number of auxiliary operators is reduced by a factor of only $1.2$, while a factor of 
about two can be obtained for lower temperatures. This is because the Pade and the Matsubara decomposition scheme share 
half of the frequencies (\ref{paramscheme3}) and the amplitudes (\ref{paramscheme2}). 
Moreover, the similarity of results and the small gain in numerical efficiency points out 
the quality of our truncation scheme (see App.\ \ref{appB}).

\subsection{Transport properties of junction CON}
\label{consec}

\begin{figure}
\begin{tabular}{l}
\hspace{-0.5cm}(a) \\
\resizebox{\newwidth}{\newheight}{
\includegraphics{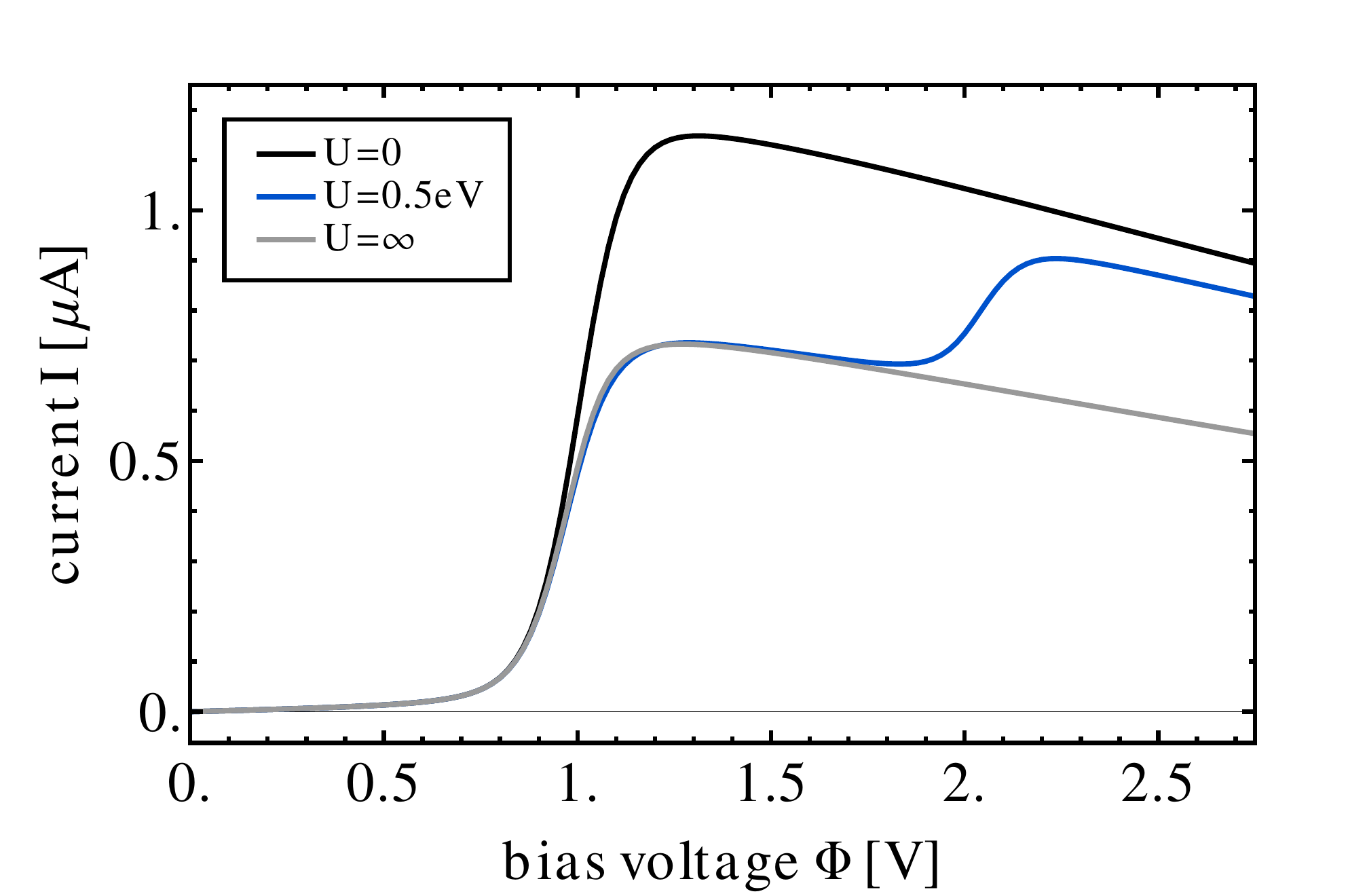}
}
\\
\hspace{-0.5cm}(b) \\
\resizebox{\newwidth}{\newheight}{
\includegraphics{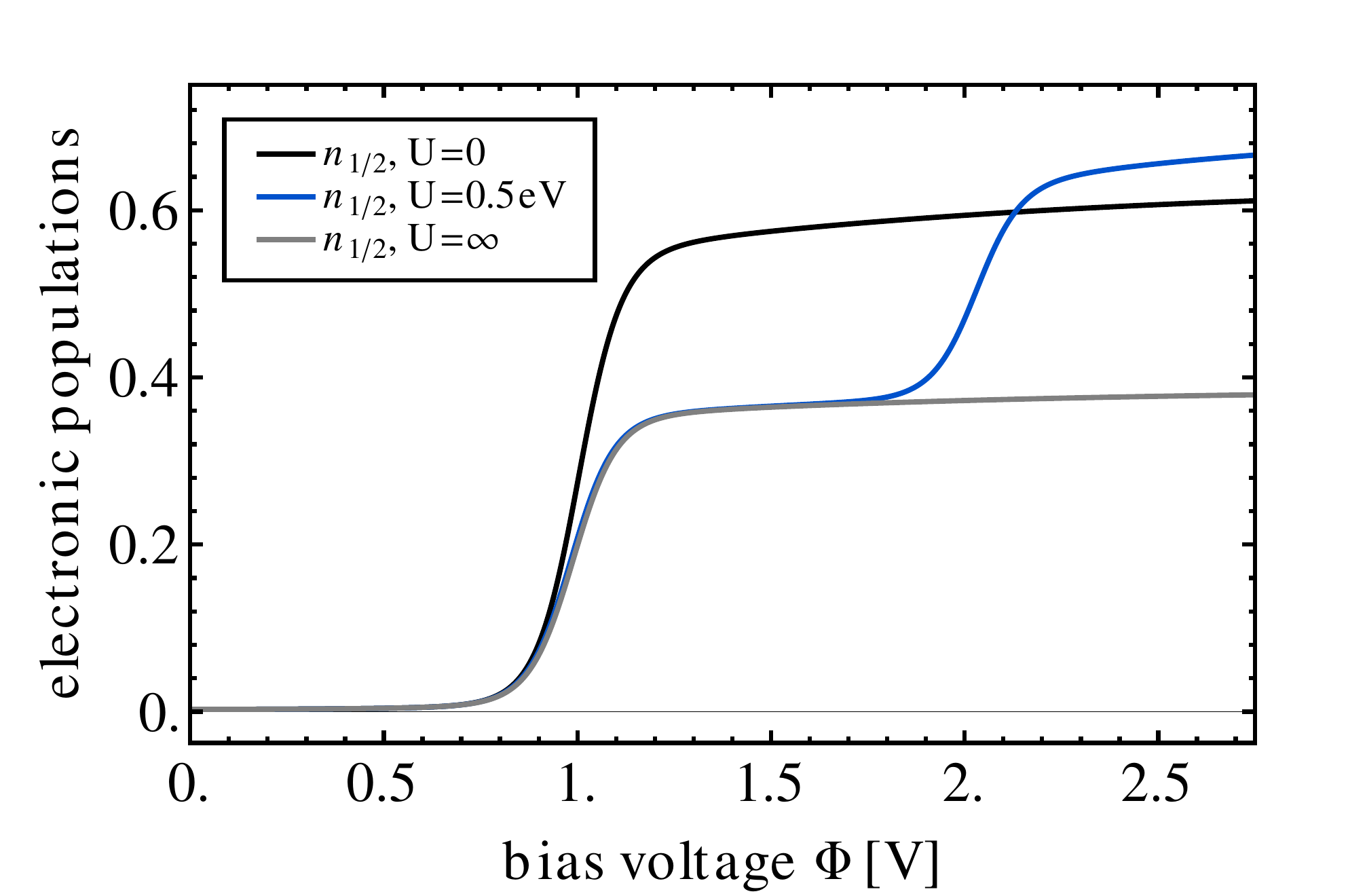}
}
\end{tabular}
\caption{(Color online)\label{modelB6current} Current-voltage and electronic population characteristics  
of junction CON. Interference effects and decoherence due to electron-electron interactions is 
less pronounced in this system. 
}
\end{figure}

To complete our discussion, we  discuss the transport properties of the complementary model system CON. 
The corresponding current-voltage and population characteristics are shown in Fig.\ \ref{modelB6current} (blue lines). 
For comparison, we also show the transport characteristics of this junction if electron-electron interactions are 
neglected (black lines) and for the limit $U\rightarrow\infty$ (gray lines). All three scenarios exhibit NDR in the 
resonant transport regime. In contrast to junction DES, the NDR is not very pronounced and not related to decoherence 
effects but rather a direct consequence of the finite band width $\gamma$ which results in a reduced overlap of the 
conduction bands at $\epsilon\approx\epsilon_{1/2}$ and $\epsilon\approx\epsilon_{1/2}+U$ and, thus, in reduced 
current levels if the bias voltage is increased. Note that this mechanism for NDR is counterbalanced in junction DES, 
because destructive interference effects become simultaneously less pronounced.

Fig.\ \ref{modelB6current}a also shows that 
the current level of junction CON is suppressed for bias voltages $2\epsilon_{1/2}<e\Phi<2(\epsilon_{1/2}+U)$. 
In this regime of 
bias voltages, the junction can be populated via two channels but depopulated only via a single channel. The 
corresponding current can thus be estimated as $2\Gamma^{2}/(2\Gamma+\Gamma)$, corresponding to a reduction of the 
current by a factor of $2/3$. This blocking of transport channels is also reflected in the population of the 
electronic levels $1$ and $2$, where the aforementioned asymmetry in the population and depopulation dynamics 
of the junction results in an average population of $\simeq1/3$. At higher bias voltages, both levels are populated 
simultaneously and the  average population increases. It exceeds $0.5$ due to the finite band width $\gamma$.

In comparison to junction DES, interference and lead induced inter-state coupling effects play no role for the transport properties 
of junction CON. The population of the two eigenstates is almost identical, $n_{1}\approx n_{2}$, and the coherence 
$\sigma_{10,01}$ takes very low values ($\vert\sigma_{10,01}\vert\lesssim10^{-3}$, data not shown). This seems 
counterintuitive at first, since the quasidegeneracy of the eigenstates and the symmetry of the coupling strengths 
$\nu_{K,m}$ implies strong constructive interference effects. Such effects indeed double the current 
level of the system in the non-resonant transport regime \cite{Hartle2012}. At higher bias 
voltages, however, an antiresonance at $\epsilon=(\epsilon_{1}+\epsilon_{2})/2$ cancels the effect of constructive 
interference effects (cf.\ Ref.\ \onlinecite{Hartle2012} for a more detailed discussion). This behavior can also be 
understood in terms of the weak coupling between the dots (cf.\ Fig.\ \ref{LinConduct}), which effectively factorizes 
the corresponding population dynamics. As a result, electron-electron interactions have only an electrostatic 
influence in this system. A similar behavior is obtained if $\vert\epsilon_{2}-\epsilon_{1}\vert>\Gamma$ (even for junction DES).

According to this analysis, junctions DES and CON can be viewed as two extreme cases, 
where decoherence due to electron-electron interactions is fully developed and where it plays no role, respectively. 
It is thus interesting to note that the BM scheme is also capable of describing resonant transport through  
junction CON both qualitatively and quantitatively (except for the broadening due to the coupling to the electrodes, 
data not shown).

\section{Conclusion}

We have investigated the electrical transport properties of an interacting double quantum dot system 
with quasidegenerate electronic 
states, focusing on decoherence phenomena due to electron-electron interactions and employing a modified version 
of the numerically exact HQME approach of Jin \emph{et al.}\ \cite{Jin2008}. The HQME approach allowed us to validate 
our results with respect to higher-order ($\geq2$) effects in a systematic, numerically exact way. This is crucial in the present 
context, as the most important effects are associated with sequential tunneling \cite{Hettler2007} while, concurrently, 
a strict second order expansion yields unphysical results in some parameter regimes. In addition, a comparison of the HQME and the 
BM scheme elucidated strong inter-state coupling effects that originate in these systems from the energy dependence of 
the tunneling efficiency between the dots and the electrodes. We find that interaction-induced 
decoherence gives rise to pronounced negative differential resistance \cite{Wunsch2005,Hettler2007,Trocha2009}, 
an enhanced broadening of the corresponding electronic signatures and an inversion of the electronic population 
of the junction (cf.\ Fig.\ \ref{model6band}). We found that these phenomena can be even more pronounced for 
intermediate asymmetries in the coupling to the electrodes. An important experimental signature which distinguishes 
decoherence-induced NDR  from other mechanisms, for example a centrally localized or a blocking state, is the 
symmetry with respect to a gate voltage around the charge symmetric point (compare Figs.\ \ref{maps}a and \ref{maps3}).

The decoherence mechanism that underlies these effects is based on the blocking of (doubly occupied) states. The 
principal effects are thus only weakly dependent on the electron-electron interaction strength and enter only via the 
relative position of resonances as well as the energy dependence of the tunneling efficiency between the 
dots and the electrodes. The enhanced broadening of electronic signatures in the transport 
characteristics and the inversion of the electronic population can be solely attributed to the effect of the latter and 
the associated inter-state coupling while the shape of the associated current peaks or the appearance of negative differential 
resistance is also strongly modified. 
As a consequence, the corresponding conductance map shows resonance lines that can be disconnected, bent and 
even smeared out (compare Figs.\ \ref{maps}a and \ref{maps2}). These results have been further corroborated by 
studying different functional forms of the tunneling efficiency between the dots and the electrodes using NCA. 
While the BM and the NCA method capture but also miss some of the qualitative aspects, they fail to describe the 
transport characteristics of these systems on a quantitative level. 
Note that, according to the results of Pedersen \emph{et al.}\ \cite{Hettler2007}, one can expect that the decoherence 
effects described in this work are even more pronounced for the spinful case.

\section*{Acknowledgment}

We thank J.\ Okamoto, M.\ Kulkarni, 
M.\ Schiro, A.\ Croy, C.\ Schinabeck and M.\ Thoss for fruitful and inspiring discussions. 
RH is grateful to the Alexander von Humboldt Foundation for the award of a Feodor-Lynen research fellowship 
and GC to the Yad Hanadiv--Rothschild Foundation for the award of a Rothschild Postdoctoral Fellowship. 
This work has also been supported by the National Science Foundation (NSF DMR-1006282 and NSF CHE-1213247).

\appendix 

\section{Numerical solution of the hierarchical equations of motion}
\label{appB}

The hierarchical equations of motion (\ref{hierarcheom}) are numerically integrated using the Euler method. Thereby, 
the initial conditions are set to the state where the system is initially unpopulated. Other initial conditions, 
associated with the singly occupied and doubly occupied system, have also been considered. The steady state properties 
studied in this article did not show any dependence on the choice of the initial states. In this context it should 
be noted that all auxiliary operators are identical to zero at $t=0$. This can be inferred from the definitions 
(\ref{defauxop2}) and (\ref{defauxop}) and corresponds to the choice that the system S and the electrodes L and R 
are initially uncorrelated.

The numerical effort in the evaluation of the hierarchical of equations of motion (\ref{hierarcheom}) 
can be reduced, using that 
\begin{eqnarray}
\label{appBrel}
 \left( \sigma^{(\alpha)}_{j_{1}..j_{\alpha}}(t) \right)^{\dagger} &=& 
 (-1)^{\sum_{\beta=1..\alpha} \beta} \sigma^{(\alpha)}_{\overline{j}_{1}..\overline{j}_{\alpha}}(t), 
\end{eqnarray}
where $\overline{j}_{1}=(K_{\beta},m_{\beta},\overline{s}_{\beta},p_{\beta})$, and 
\begin{eqnarray}
\label{appBrel2}
 B_{j_{1}} B_{j_{2}} &=& - B_{j_{2}} B_{j_{1}} .
\end{eqnarray}
The latter identity implies that the auxiliary operators $\sigma^{(\alpha)}_{j_{1}..j_{\alpha}}(t)$ are identical to 
zero if two of the superindices $j_{1}$ -- $j_{\alpha}$ coincide \cite{Jin2008}. It should be noted at this point that 
there is, in general, an infinite number of superindices $j_{\beta}$ and, accordingly, of auxiliary operators such that 
the hierarchy of equations of motion (\ref{hierarcheom}) does in general not terminate automatically. 
Only in specific limits, in particular the non-interacting limit ($U=0$) or for $U\rightarrow\infty$, 
the hierarchy of equations of motion (\ref{hierarcheom}) truncates automatically at the second tier \cite{Jin2008,Jin2010}. 
Respecting  the relations (\ref{appBrel}) and (\ref{appBrel2}), the number of auxiliary operators is given by $2^{N-1}$, 
where $N$ denotes the number of different superindices $j_{\beta}$.

The above result for the number of auxiliary operators only applies if all tiers of the hierarchy (\ref{hierarcheom}) 
are taken into account. In practice, however, it is truncated at a finite tier $\alpha$. 
This reduces the number of auxiliary operators considerably to $\sum_{\beta=1..\alpha} \frac{N!}{2\beta!(N-\beta)!}\sim N^{\alpha}/\alpha!$. 
The number of auxiliary operators can be reduced even further, because 
the frequencies $\omega_{K,p}^{s}$ and the amplitudes $\eta^{s}_{K,p}$ typically vary by orders of magnitude. 
The amplitudes $\eta^{s}_{K,p}$ enter the definition of the auxiliary operators as 
$\prod_{\beta=1..\alpha}\eta^{s_{\beta}}_{K_{\beta},p_{\beta}}$ (cf.\ Eq.\ (\ref{defauxop})), while the real parts 
of the frequencies $\text{Re}[\omega^{\pm}_{K,p}]$ determine the time scales, where the dynamics of the system influences 
these quantities. The relative importance of the auxiliary operators $\sigma^{(\alpha)}_{j_{1}..j_{\alpha}}(t)$ can thus 
be assessed by the dimensionless amplitude $\prod_{\beta=1..\alpha}\vert\eta^{s_{\beta}}_{K_{\beta},p_{\beta}}\vert / 
\text{Re}[\omega^{s_{\beta}}_{K_{\beta},p_{\beta}}]$. Furthermore, the importance of the respective tier level $\alpha$ can 
be estimated by the prefactors $\nu_{K_{\alpha},m_{\alpha}}\nu_{K_{\alpha},n_{\alpha}}/\gamma$ and the sum of the frequencies 
$\sum_{\beta=1..\alpha} \omega_{K_{\beta},p_{\beta}}^{s_{\beta}} $, which link auxiliary operators of different tiers in 
Eqs.\ (\ref{hierarcheom}), in particular when the system is close to a stationary state. Therefore, in practical calculations, 
we neglect all auxiliary operators in the second and higher tier that have a dimensionless amplitude $\left\vert 
\gamma^{1-\alpha} \left( \prod_{\beta=1..\alpha-1}\nu_{K_{\beta},n_{\beta}}^{2} / \sum_{\beta'=1..\beta} 
\text{Re}[\omega^{s_{\beta'}}_{K_{\beta'},p_{\beta'}}] \right) \cdot \left(\prod_{\beta=1..\alpha} \eta^{s_{\beta}}_{K_{\beta},p_{\beta}} / 
\text{Re}[\omega^{s_{\beta}}_{K_{\beta},p_{\beta}}] \right) \right\vert$ that is smaller than a threshold value $A_{\text{th}}$. 
This value is reduced and, concurrently, the number of auxiliary operators in the first tier 
is increased until converged results are obtained. 
Note that, due to this choice of the amplitudes, it is not relevant whether the factor $\pi$, which appears in Eqs.\ (\ref{paramscheme2}), 
is included in the amplitudes $\eta_{K,p}^{s}$ or in the prefactors $\nu_{K_{\alpha},m_{\alpha}}\nu_{K_{\alpha},n_{\alpha}}/\gamma$.

Using this truncation scheme, the numerical effort can thus be reduced to a feasible level, as is further discussed  in 
Sec.\ \ref{numerisec}. The numerical effort scales as $\sim N^{\alpha-1}/\alpha!$ considering that the amplitude condition 
cuts out a hypersurface of the total index space. For the specific parametrization scheme employed in Eq.\ (\ref{paramscheme}) 
this statement can be formulated more explicitly. As 
\begin{eqnarray}
 \vert\eta_{K,p}^{s}\vert/\text{Re}[\omega_{K,p}^{s}] \sim \frac{1}{ 2p-1 } 
\frac{1}{(2p-1)^{2} - (\frac{\gamma}{k_{\text{B}}T})^{2} } 
\end{eqnarray} 
an upper bound for the width and the radius of the corresponding hypersurface may be estimated as $\sim 1/A_{\text{th}}$ 
and $\sim \gamma/(k_{\text{B}}T)$, respectively. Accordingly, the number of auxiliary quantities at each tier $\alpha$ scales 
as $\gamma^{\alpha-1}/((k_{\text{B}}T)^{\alpha-1}A_{\text{th}}\alpha!)$. Finally, it should be noted that the density and all 
auxiliary operators include $2^{N_{\text{el}}}$ matrix elements.

\end{document}